%% file: Vortex_Structure_arXiv.tex
\documentclass[a4paper,aps,prd,floatfix,reprint,twoside,twocolumn,preprintnumbers,showkeys,showpacs,final]{revtex4-2}
\usepackage[utf8]{inputenc}
\usepackage{amsmath, amssymb}
\usepackage{siunitx}
\usepackage{graphicx}
\usepackage{dcolumn}
\usepackage[caption=false]{subfig}
\usepackage{xcolor}
\usepackage{listings}
\usepackage{enumitem}
\usepackage{booktabs}
\usepackage{xr-hyper}

\usepackage{hyperref} 

\usepackage{environ} \usepackage{tikz}
\usetikzlibrary{arrows,decorations.markings}
\usetikzlibrary[arrows.meta,bending]
\usetikzlibrary{positioning}

\usepackage{media9}

\makeatletter \newsavebox{\measure@tikzpicture}
\NewEnviron{scaletikzpicturetowidth}[1]{%
  \def\tikz@width{#1}%
  \begin{lrbox}{\measure@tikzpicture}%
    \BODY
  \end{lrbox}%
  \pgfmathparse{#1/\wd\measure@tikzpicture}%
  \BODY } \makeatother

\newcommand{\mc}[1]{\multicolumn{1}{c}{#1}}

\DeclareMathOperator{\Tr}{Tr}

\newcommand{\vb}{\textbf}

\begin{document}
\preprint{ADP-23-04/T1213}
\title{Centre vortex structure in the presence of dynamical fermions} \author{James C.
  Biddle} \author{Waseem Kamleh} \author{Derek B. Leinweber} \affiliation{Centre
  for the Subatomic Structure of Matter, Department of Physics, The University
  of Adelaide, SA 5005, Australia}
\begin{abstract}
  An analysis of the geometry and structure of centre vortices in the presence
  of dynamical fermions is performed. A variety of metrics are used to measure
  the matrix structure of the vortex-modified gauge fields. Visualisations of
  centre vortices are presented and percolating clusters are identified. The
  size of secondary vortex clusters is analysed, with substantial differences
  observed between the pure Yang-Mills and dynamical fermion case. Vortex fields
  are represented as directed graphs, with branching points acting as the
  vertices. This representation leads to a novel picture of vortex branching as
  a binomial process. These results elucidate the change in the centre vortex
  vacuum induced by the introduction of dynamical fermions.
\end{abstract}
\maketitle

\section{Introduction}
There is now a wealth of literature exploring the impact of centre vortices on
pure Yang-Mills gauge
theory~\cite{tHooft:1977nqb,tHooft:1979rtg,DelDebbio:1996lih,Faber:1997rp,DelDebbio:1998luz,Bertle:1999tw,Faber:1999gu,Engelhardt:1999fd,Engelhardt:1999xw,Engelhardt:2000wc,Bertle:2000qv,Langfeld:2001cz,Greensite:2003bk,Bruckmann:2003yd,Engelhardt:2003wm,Boyko:2006ic,Ilgenfritz:2007ua,Bornyakov:2007fz,OCais:2008kqh,Engelhardt:2010ft,Bowman:2010zr,OMalley:2011aa,Trewartha:2015ida,Greensite:2016pfc,Biddle:2018dtc,Spengler:2018dxt}.
These results have consistently shown that centre vortices play an important
role in the emergence of non-perturbative properties. However, there have also
been consistent discrepancies between original and vortex-only calculations.
Recent results~\cite{Biddle:2022zgw,Biddle:2022acd} have for the first time
considered centre vortices in the presence of dynamical fermions. These results
demonstrated the dramatic effect dynamical fermions have on the behaviour of
centre vortices. In contrast to prior pure Yang-Mills
studies~\cite{Trewartha:2015nna,OMalley:2011aa,Trewartha:2017ive,Langfeld:2003ev,Bowman:2010zr,Biddle:2018dtc,Bowman:2010zr,OCais:2008kqh},
the static quark potential can be fully recreated from centre vortices
alone~\cite{Biddle:2022zgw}, and vortex removal results in complete suppression
of the infrared Landau-gauge gluon propagator~\cite{Biddle:2022acd}. In light of
these unexpected results, it is natural seek a deeper understanding of these effects by
directly analysing the structure of the vortices themselves.

In this work, we first look for changes in the bulk properties of the lattice
configurations by analysing the norms and traces of the gauge links, as well as
the values of the maximal centre gauge functional. Bulk discrepancies between
pure-gauge and dynamical ensembles may suggest where the differences in vortex
structure arise from.

We then expand upon the visualisation techniques developed in
Ref.~\cite{Biddle:2019gke} to analyse the geometric structure of centre
vortices. New developments allow us to split the vortex structure into
individual disconnected clusters. From these clusters we may examine the degree
of vortex percolation present in the vacuum.

In the supplemental material located at the end of this document, visualisations
of these centre vortex clusters are presented as interactive 3D models embedded
in the document. Instructions on viewing these models are also included therein.
Figures with a corresponding interactive model that can be found in the
supplemental material are marked as \textbf{Interactive} in the caption.
Interactive models in the supplementary material are also referenced as Fig.~S-x
in the text. A selection of preset views that highlight regions of interest is
also available.

Following cluster identification, we present a novel perspective that considers
each cluster as a directed graph of vortex branching points, with the weight of
each graph edge corresponding to the number of vortex plaquettes between branching
points. This data structure enables us to develop quantitative measures of the
size and shape of centre vortex clusters, facilitating a detailed comparison of
vortex structure between pure-gauge and dynamical QCD.

This paper is structured as follows. In Sec.~\ref{sec:CentreVortices} we detail
the centre vortex model and how centre vortices are identified on the lattice.
We then present the analysis of the bulk gauge link properties in
Sec.~\ref{sec:BulkProperties}. In Sec.~\ref{sec:Visualisations} our
visualisation conventions are introduced. In Sec.~\ref{sec:LoopID} we
discuss the cluster identification algorithm and subsequent findings. In
Sec.~\ref{sec:BPGraph} we introduce the method by which vortex clusters can be
converted to a graph, and discuss the analysis performed on these graphs.
Finally, the findings of this work are summarised in Sec.~\ref{sec:Conclusion}.

\section{Centre Vortices}\label{sec:CentreVortices}
In QCD, centre vortices are regions of a gauge field that carry flux associated
with $\mathbb{Z}_{3}$, the centre of the $SU(3)$ gauge group. $\mathbb{Z}_{3}$
consists of the three elements,
\begin{equation}
  \label{eq:Z3}
  \mathbb{Z}_3 =  \left\lbrace\exp\left(m\frac{2\pi i}{3} \right)I, ~ m = -1, 0, +1\right\rbrace\, .
\end{equation}
For the purposes of our discussion, $m$ will be referred to as the centre charge
of the vortex. On the lattice, thin centre vortices appear as closed sheets in four
dimensions, or as closed lines on three dimensional slices
of the lattice.

Centre vortices are identified on the lattice through a well-known procedure~\cite{Biddle:2019gke,Montero:1999by},
briefly summarised here. First, the configurations are rotated to maximal centre
gauge (MCG) by determining a gauge rotation, $\Omega(x)$, that maximises the
functional~\cite{Langfeld:2003ev,Trewartha:2015ida,Montero:1999by}
\begin{equation}
  \label{eq:MCGfunc}
  \Phi = \frac{1}{V\, N_\text{dim}\, n_c^2} \sum_{x, \mu}\left| \Tr U^{\Omega}_{\mu}(x) \right|^2\, .
\end{equation}
This process brings each gauge link as close as possible to one of the elements
of $\mathbb{Z}_3$. Once the ensemble has been fixed to maximal centre gauge,
each link is projected onto the nearest centre element, $U_{\mu}(x)\rightarrow Z_{\mu}(x)$, as defined by the phase
of the trace of each link. Centre vortices are then identified by the location
of non-trivial plaquettes $P_{\mu\nu} = \exp\left(m\frac{2\pi i}{3} \right)I$, in the $\mu$\nobreakdash--$\nu$ plane
with $m=\pm 1$. This process of centre projection defines the vortex-only
ensemble, $Z_{\mu}(x)$. Using these identified vortices, we also construct the
vortex-removed ensemble by computing
$R_{\mu}(x)=Z_{\mu}^{\dagger}(x)\,U_{\mu}(x)$. Hence, this procedure results in
three ensembles:
\begin{enumerate}
\item Original, untouched (UT) fields, $U_{\mu}(x)$,
\item Vortex-only (VO) fields, $Z_{\mu}(x)$,
\item Vortex-removed (VR) fields, $R_{\mu}(x)$,
\end{enumerate}

Visualisations of vortices are naturally constructed from the vortex-only
ensembles, and as such the $Z_{\mu}(x)$ fields will be of primary focus in this
work. However, the effectiveness of vortex removal is also of great interest as
it has been observed that the vortex removed ensembles also vary in behaviour
depending on the presence or absence of dynamical fermions~\cite{Biddle:2022acd,Biddle:2022zgw}.

For this work, we continue the analysis performed in our
previous work~\cite{Biddle:2022acd,Biddle:2022zgw} and make use of three original (UT)
ensembles. Each ensemble has dimensions $32^3\times 64$ and is comprised of 200
lattice configurations. Two of the ensembles are $(2 + 1)$ flavour dynamical
ensembles from the PACS-CS collaboration~\cite{Aoki:2008sm}. We choose the
heaviest and lightest pion mass ensembles, with masses of $701~\si{MeV}$ and
$156~\si{MeV}$ respectively. This allows us to observe the greatest
differentiation between the dynamical ensembles. The third ensemble is pure
Yang-Mills, generated with the Iwasaki gauge action~\cite{Iwasaki:1983iya}. The
lattice spacing is tuned to be similar to that of the PACS-CS ensembles. A
summary of the lattice parameters is provided in Table~\ref{tab:LatticeParams}.

\begin{table}[tb]
  \caption{A summary of the lattice ensembles used in
    this work~\cite{Aoki:2008sm}.}
  \label{tab:LatticeParams}
  \begin{ruledtabular}
    \begin{tabular}{lcccc}
      Type & $a\, (\si{fm})$ & $\beta$ & $\kappa_{\rm u,d}$ & $m_{\pi}\, (\si{MeV})$ \\
      \hline\\
      Pure gauge & 0.100 & 2.58 & - & - \\
      Dynamical & 0.102 & 1.9 & 0.13700 & 701\\
      Dynamical & 0.093 & 1.9 & 0.13781 & 156
    \end{tabular}
  \end{ruledtabular}
\end{table}

\section{Bulk Properties}\label{sec:BulkProperties}
\begin{figure}
  \centering
  \includegraphics[width=\linewidth]{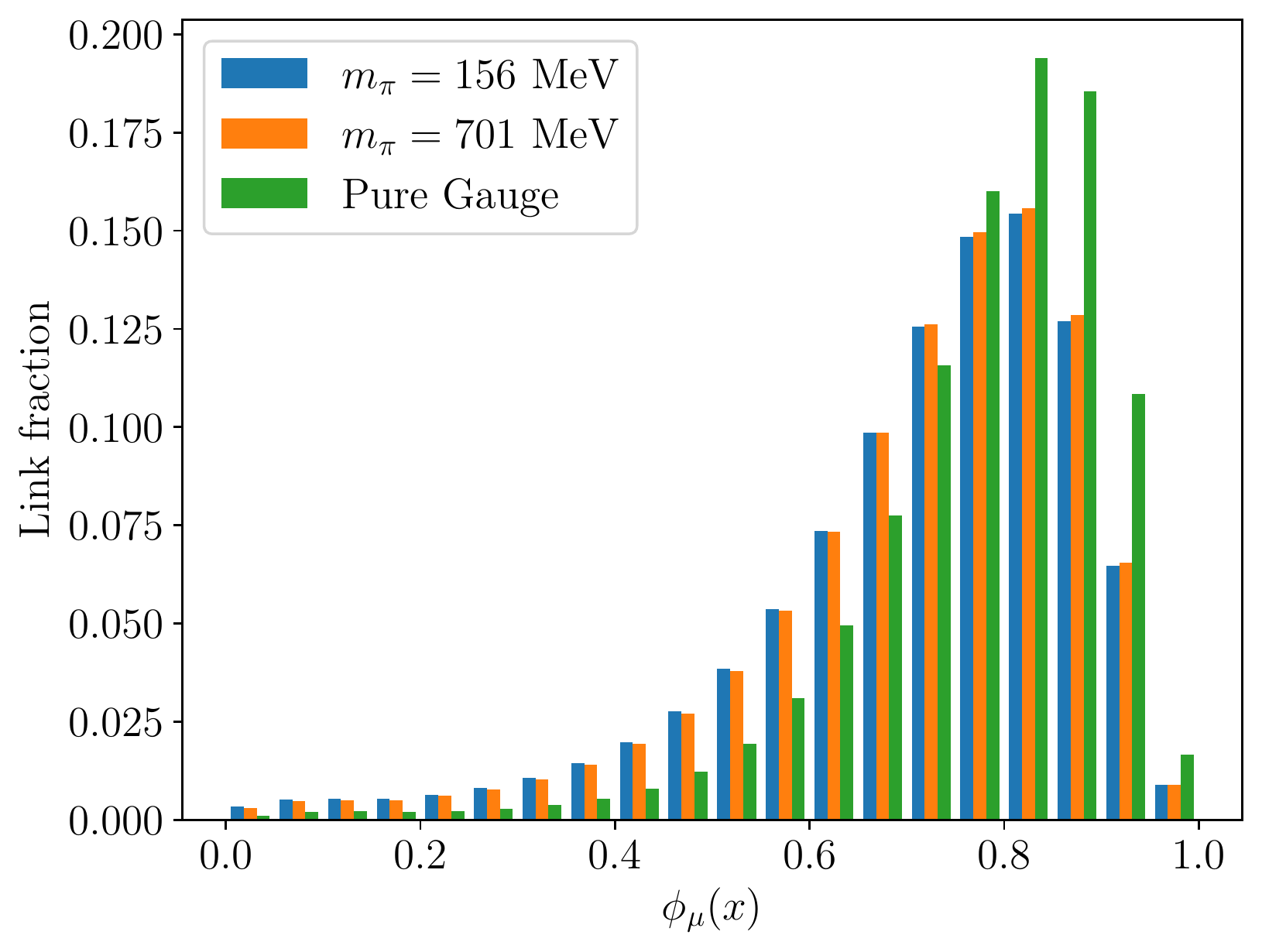}
  \caption{Distribution of the local maximal centre gauge functional,
    $R_{\mu}(x)$, as defined in Eq.~\ref{eq:localR}}
  \label{fig:Rvalues}
\end{figure}

In understanding the impact dynamical fermions have on the centre-vortex vacuum,
it is natural to first look for bulk changes in the $SU(3)$ lattice gauge fields
upon the introduction of dynamical fermions. The first measure we examine is the
distribution of the local MCG functional
\begin{equation}
  \label{eq:localR}
\phi_{\mu}(x) = \frac{1}{n_{c}^2} \left| \Tr U^{\Omega}_{\mu}(x) \right|^2
\end{equation}
defined such that the total MCG functional given in Eq.~\eqref{eq:MCGfunc} can be written as
\begin{equation}
  \label{eq:2}
  \Phi=\frac{1}{V\,N_{\rm dim}}\sum_{x,\,\mu}\phi_{\mu}(x)
\end{equation}
The distribution of $R_{\mu}(x)$ values is presented for the untouched ensembles in Fig.~\ref{fig:Rvalues}.

We observe that the pure gauge ensemble achieves a typically larger value of $\phi_{\mu}(x)$, indicating that the links have been brought closer to the centre of $SU(3)$. The two dynamical ensembles follow each other rather closely, although the heavier pion mass appears to achieve slightly larger $\Phi$ values than its lighter counterpart. It should be noted however that larger values of $\phi_{\mu}(x)$ do not necessarily indicate that the MCG algorithm has performed better on these ensembles. As was determined in Refs.~\cite{OCais:2008kqh,Kovacs:1999st,Bornyakov:2000ig}, there are a number of methods that can be used to increase the typical values of $\phi_{\mu}(x)$ obtained from maximal centre gauge. However, these methods do not necessarily improve the vortex-finding abilities of the procedure and in some cases actually degrade the vortex-finding performance. As such, it should be understood that the results presented in Fig.~\ref{fig:Rvalues} are simply showing a noticeable change in behaviour as we transition from pure gauge to dynamical ensembles, and not necessarily a worsening of vortex identification.

\begin{figure}
  \centering
  \includegraphics[width=\linewidth]{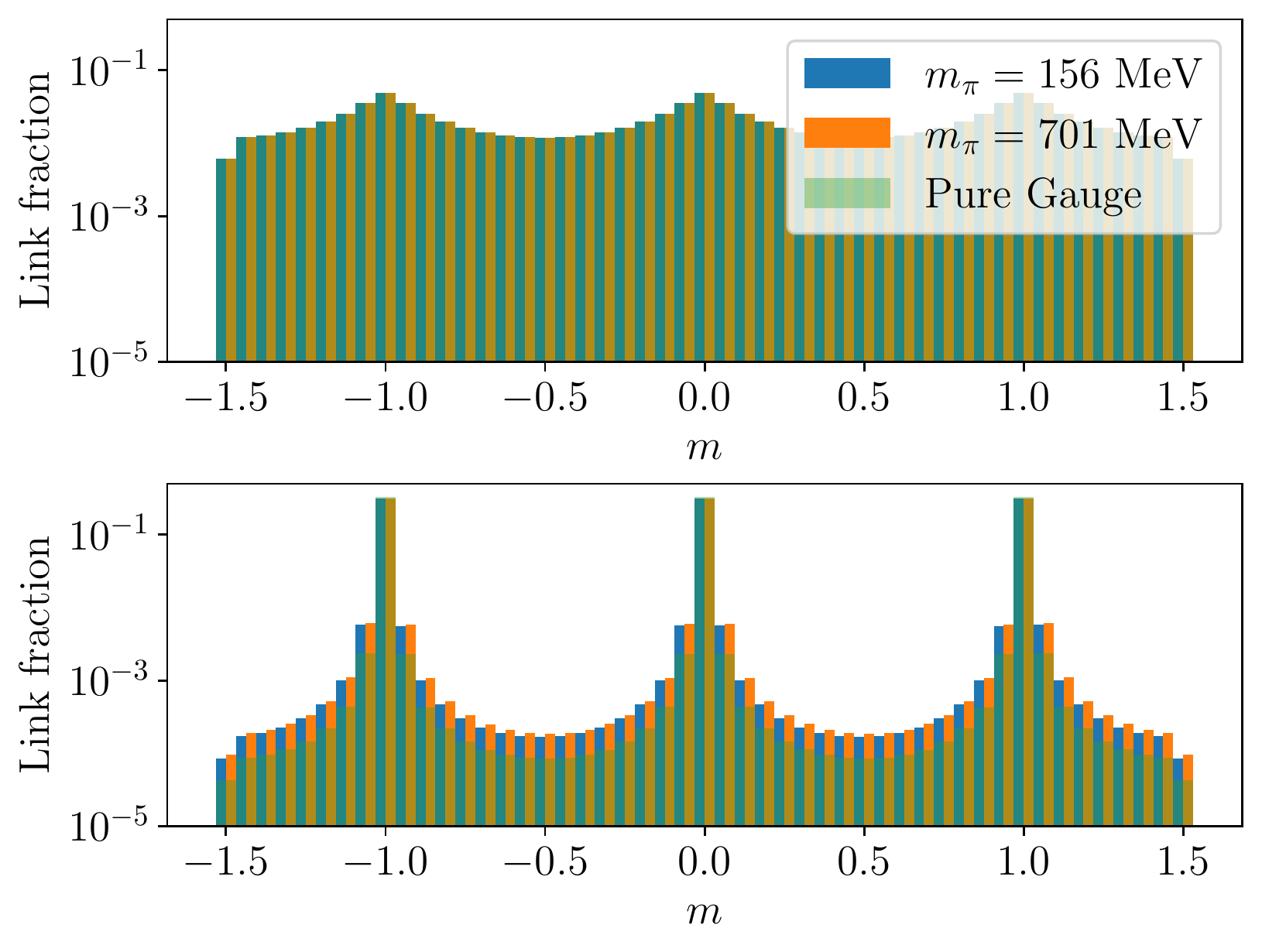}
  \caption{Distribution of trace phases before ({\bf top}) and after ({\bf
      bottom}) fixing to MCG. We plot the bins for the dynamical ensembles side-by-side as they are similar to one another, with the pure gauge results overlayed.}
  \label{fig:MCGphases}
\end{figure}

Next, we wish to compare the distribution of the trace phases, $\arg \left( \Tr U_{\mu}(x) \right)$, from each ensemble both before and after fixing to maximal centre gauge. These results are presented in Fig.~\ref{fig:MCGphases}. As intended, the phases are tightly packed about the three centre values after fixing to maximal centre gauge. However, the pure-gauge results are distributed slightly closer to the centre elements than the dynamical ensembles.

\begin{figure}
  \centering
  \includegraphics[width=\linewidth]{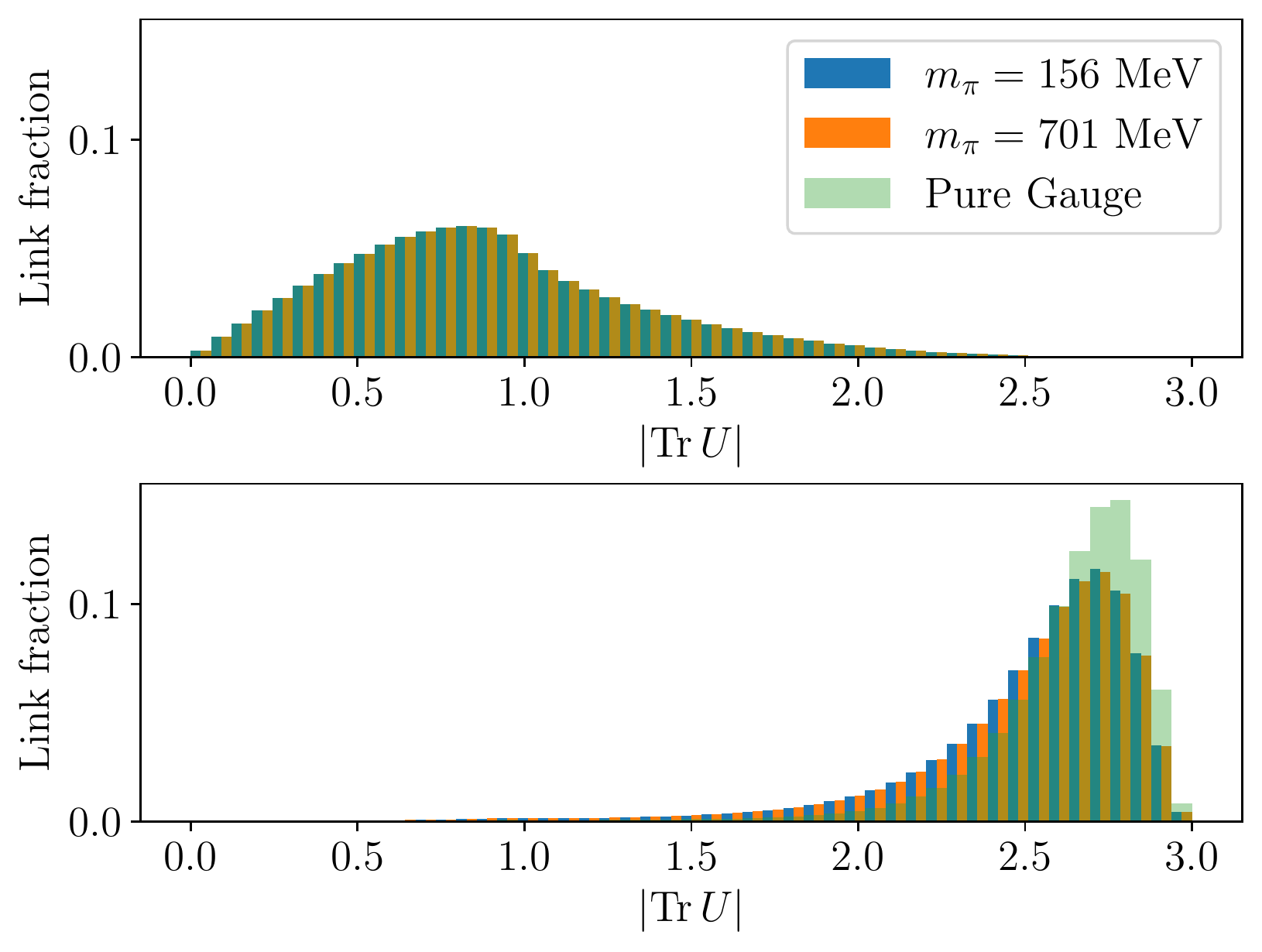}
  \caption{Distribution of trace magnitudes before ({\bf top}) and after ({\bf
      bottom}) fixing to MCG.}
  \label{fig:MCGmags}
\end{figure}

In conjunction with the trace phases, we can also look at the magnitude of the traces, $|\Tr U_{\mu}(x)|$. These values are presented in Fig.~\ref{fig:MCGmags}. Note that a centre element will have $|\Tr U_{\mu}(x)| = 3$. MCG then clearly serves to not only bring the phases close to that of a centre element, but also the magnitude. However, the effect on the magnitude is less than that on the phase. This suggests that there is still significant off-diagonal strength in the original ensembles after fixing to maximal centre gauge. Again, the pure gauge values are distributed closer to the centre value of 3 when compared with the dynamical results.

\begin{figure}
  \centering
  \includegraphics[width=\linewidth]{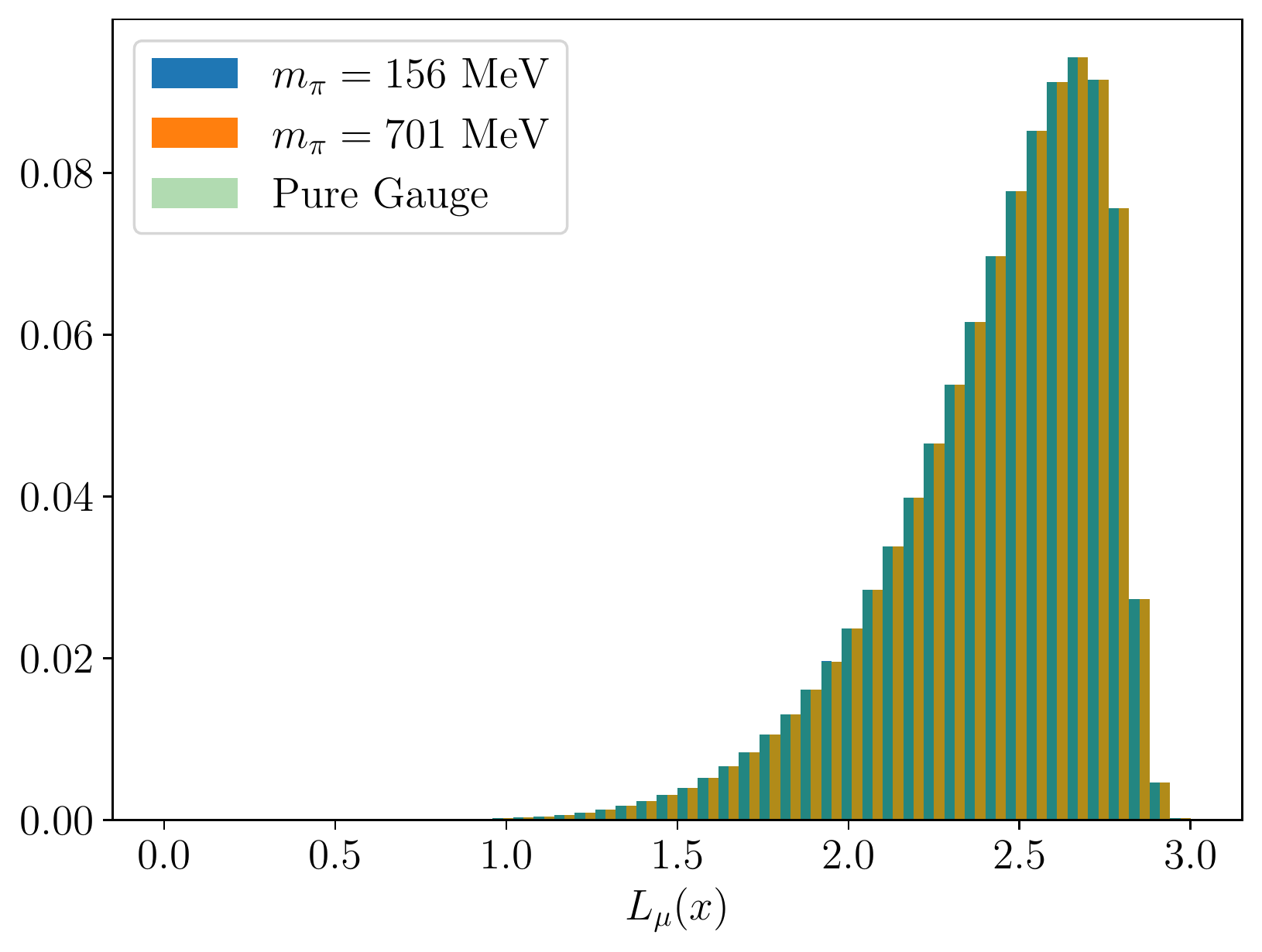}
  \caption{The $L_{\mu}(x)$ norm calculated prior to fixing to MCG.}
  \label{fig:UTnorm}
\end{figure}

\begin{figure}
  \centering
  \includegraphics[width=\linewidth]{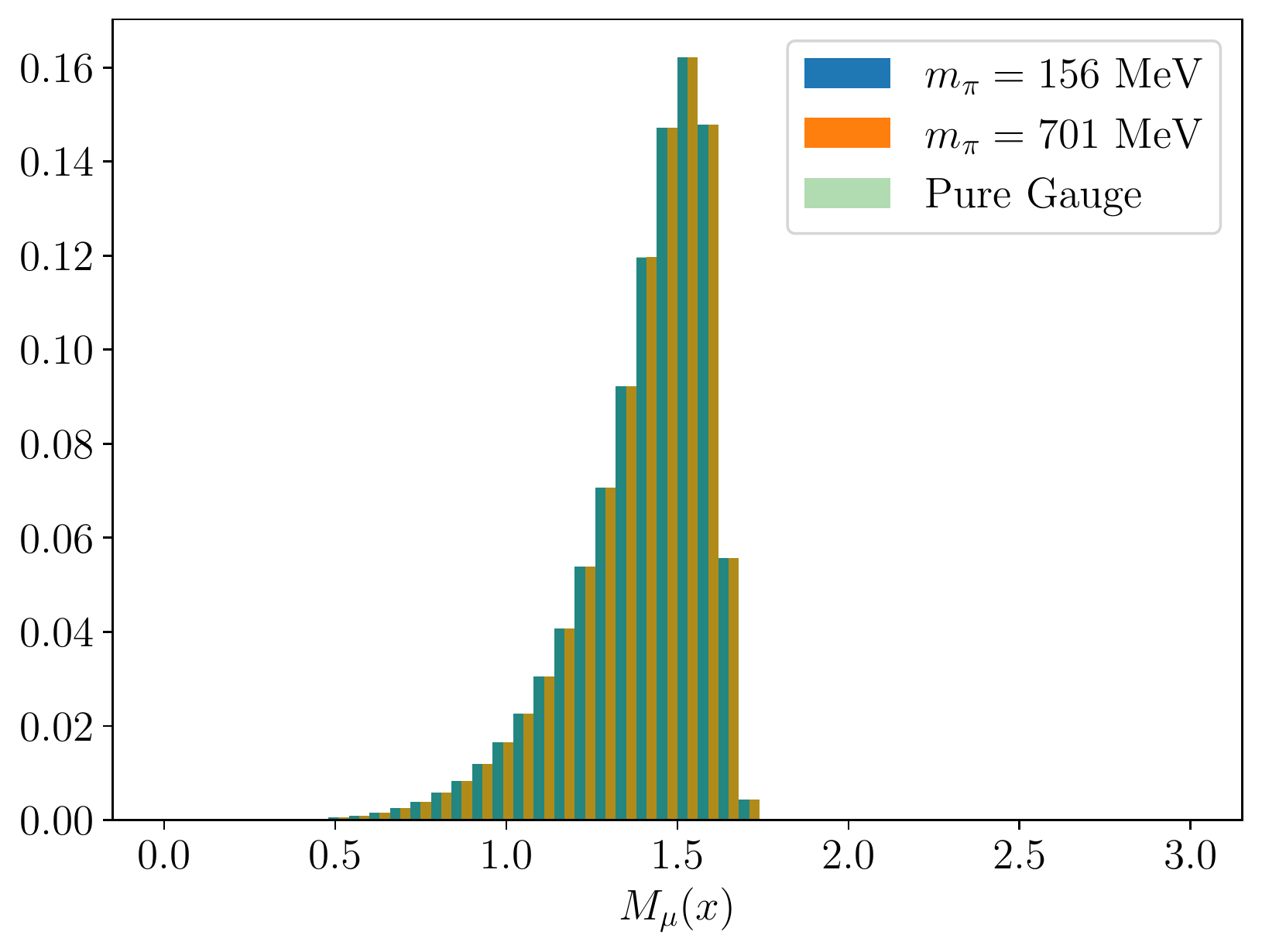}
  \caption{The $M_{\mu}(x)$ norm calculated prior to fixing to MCG.}
  \label{fig:UToffdiag}
\end{figure}

\begin{figure}
  \centering
  \includegraphics[width=\linewidth]{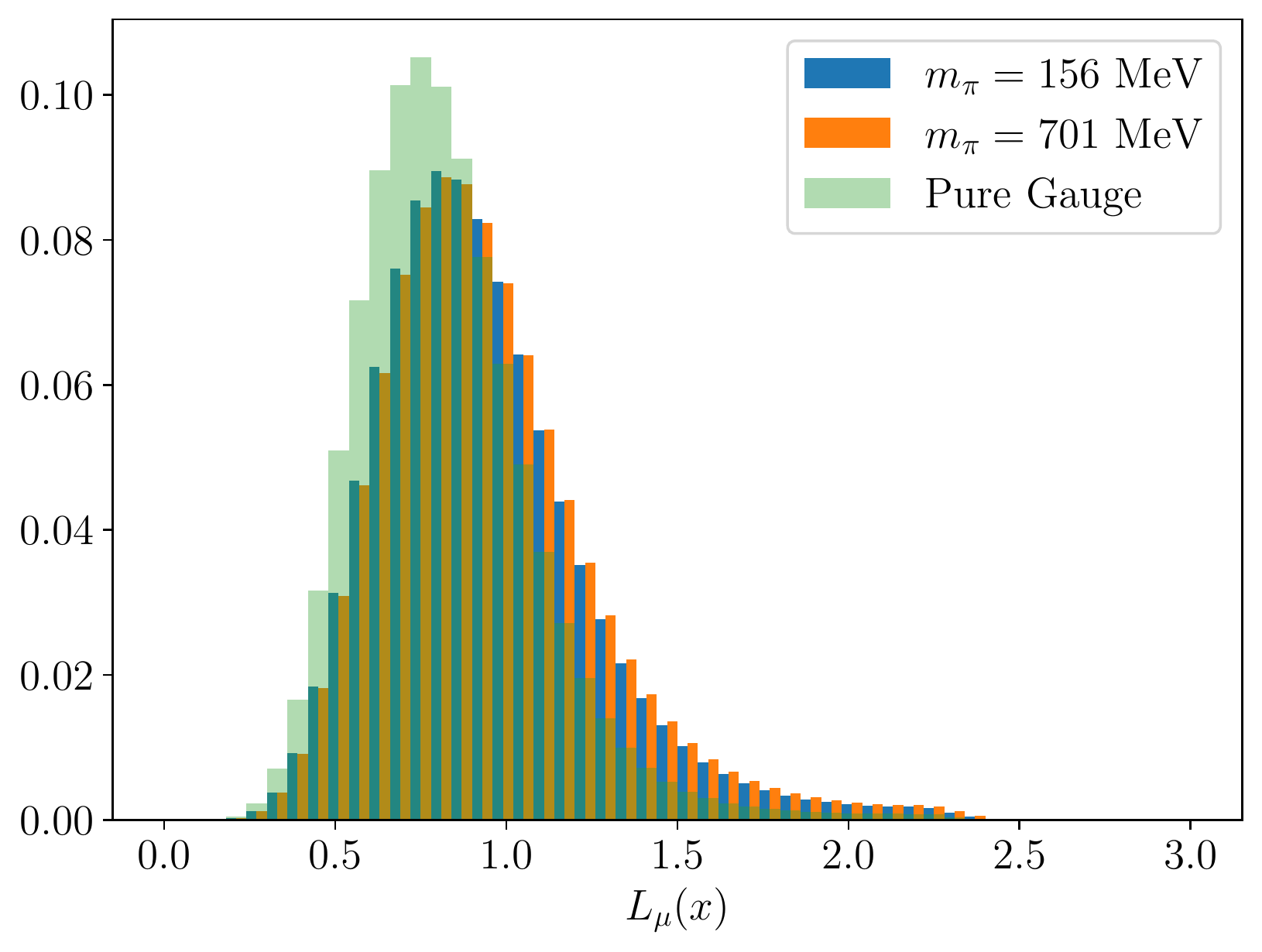}
  \caption{The $L_{\mu}(x)$ norm calculated on the VR ensembles. Here we see the
    change in behaviour after the introduction of dynamical fermions.}
  \label{fig:VRnorm}
\end{figure}

\begin{figure}
  \centering
  \includegraphics[width=\linewidth]{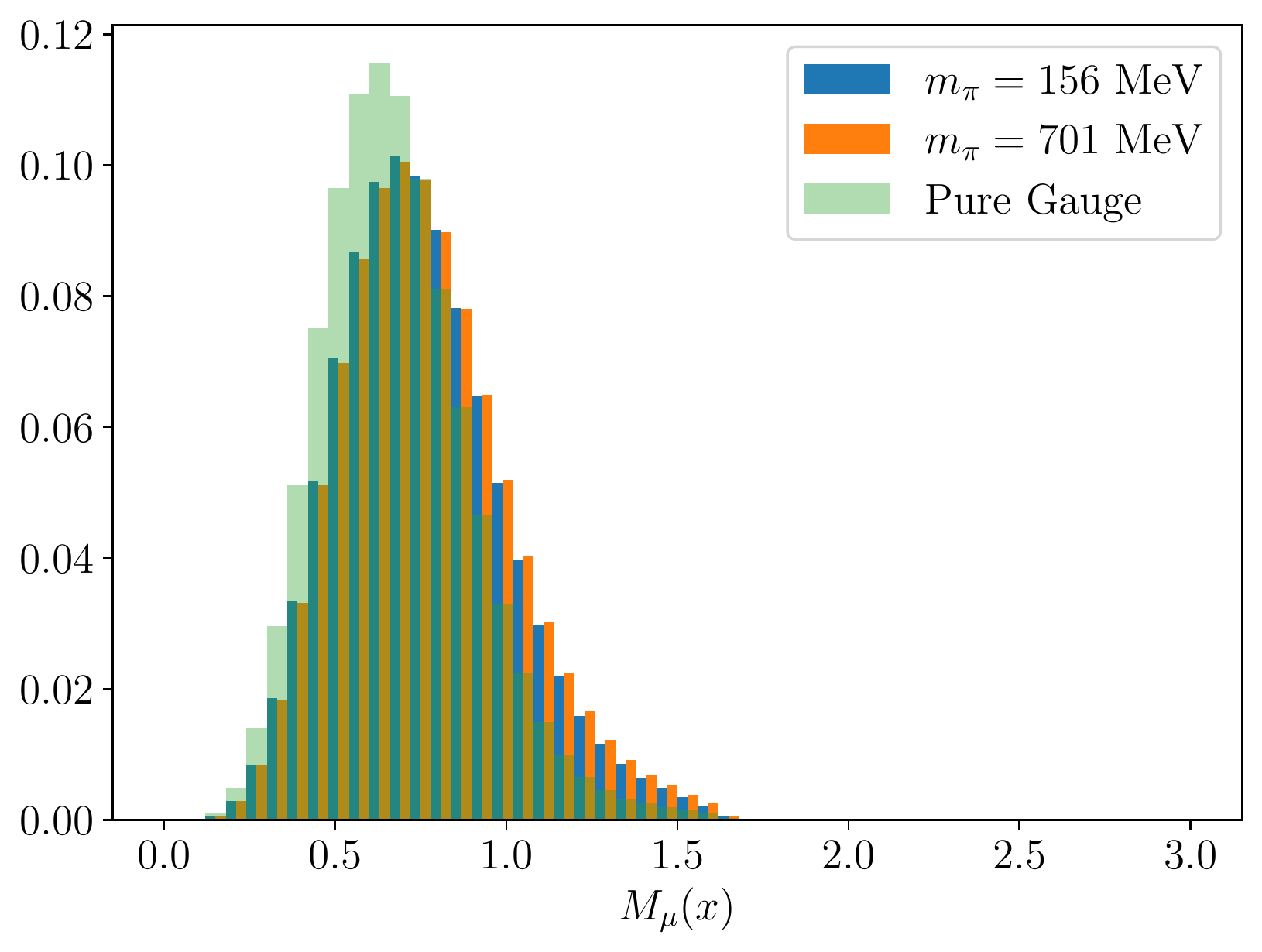}
  \caption{The $M_{\mu}(x)$ norm calculated on the VR ensembles. A trend similar
    to that seen in Fig.~\ref{fig:VRnorm} is observed.}
  \label{fig:VRoffdiag}
\end{figure}

The next bulk measures we examine are two matrix norms designed to determine the residual off-diagonal strength present in the vortex-removed fields in MCG. The norms are
\begin{equation}
  \label{eq:norm1}
  L_{\mu}(x) = \left( \sum_{i,\,j}\left| U_{\mu}^{ij}(x) - \delta_{ij} \right|^2 \right)^{\frac{1}{2}}
\end{equation}
and
\begin{equation}
  \label{eq:norm2}
  M_{\mu}(x) = \left( \sum_{\substack{i,\,j\\ i\ne j}} \left| U_{\mu}^{ij}(x) \right|^2 \right)^{\frac{1}{2}}
\end{equation}
We find for the untouched configurations that the results for both norms are
identical across all ensembles, as shown in Figs.~\ref{fig:UTnorm} and
\ref{fig:UToffdiag}. However, after vortex removal we notice that differences
appear in both norms. The results for $L_\mu(x)$ and $M_\mu(x)$ on the vortex
removed ensembles are shown in Fig.~\ref{fig:VRnorm} and
Fig.~\ref{fig:VRoffdiag} respectively.

We observe that the dynamical ensembles retain a greater proportion of their
off-diagonal strength. This is interesting, as it has been shown in
Ref.~\cite{Biddle:2022acd} that vortex removal results in a more significant
loss of infrared strength in the Landau-gauge gluon propagator when dynamical
fermions are present. This indicates that the residual strength as measured by
our norms in MCG does not coincide with enhancement as measured via the
Landau-gauge gluon propagator.

These measures indicate that there is a substantial difference in behaviour
between the pure-gauge and dynamical ensembles when considering their MCG matrix
substructure. Both the trace phases and magnitudes are further from the centre
elements and the dynamical ensembles retain more off-diagonal
strength.

\section{Visualisations}\label{sec:Visualisations}
Motivated by the difference in the bulk structure of the gauge fields in maximal
centre gauge, we now wish to look more closely at the fine-grained structure of
the vortex vacuum. We do this by extending the visualisation techniques
first developed in Ref.~\cite{Biddle:2019gke}. Given that vortices are
associated with non-trivial plaquettes, vortices themselves exist on the dual
lattice. Hence, for a vortex-only ensemble we write the plaquette
as~\cite{Engelhardt:2003wm,Spengler:2018dxt}
\begin{equation}
  \label{eq:VortexPlaq}
  P_{\mu\nu}(x)=\exp \left( \frac{\pi i}{3}\,\epsilon_{\mu\nu\kappa\lambda}\,m_{\kappa\lambda}(\bar{x}) \right)\,,
\end{equation}
where $m_{\kappa\lambda}(\bar{x})\in \left\lbrace -1,\,0,\,1 \right\rbrace$ defines the directed vortex charge orthogonal to the plaquette and based at $\bar{x}=x + \frac{a}{2}(\hat{\mu}+\hat{\nu}-\hat{\kappa}-\hat{\lambda})$. Note also that $m_{\kappa\lambda}(\bar{x})$ is anti-symmetric under index permutation, such that there is a natural association between the sign of $m$ and the vortex orientation.

\begin{figure}
  \centering
  \includegraphics[width=\linewidth]{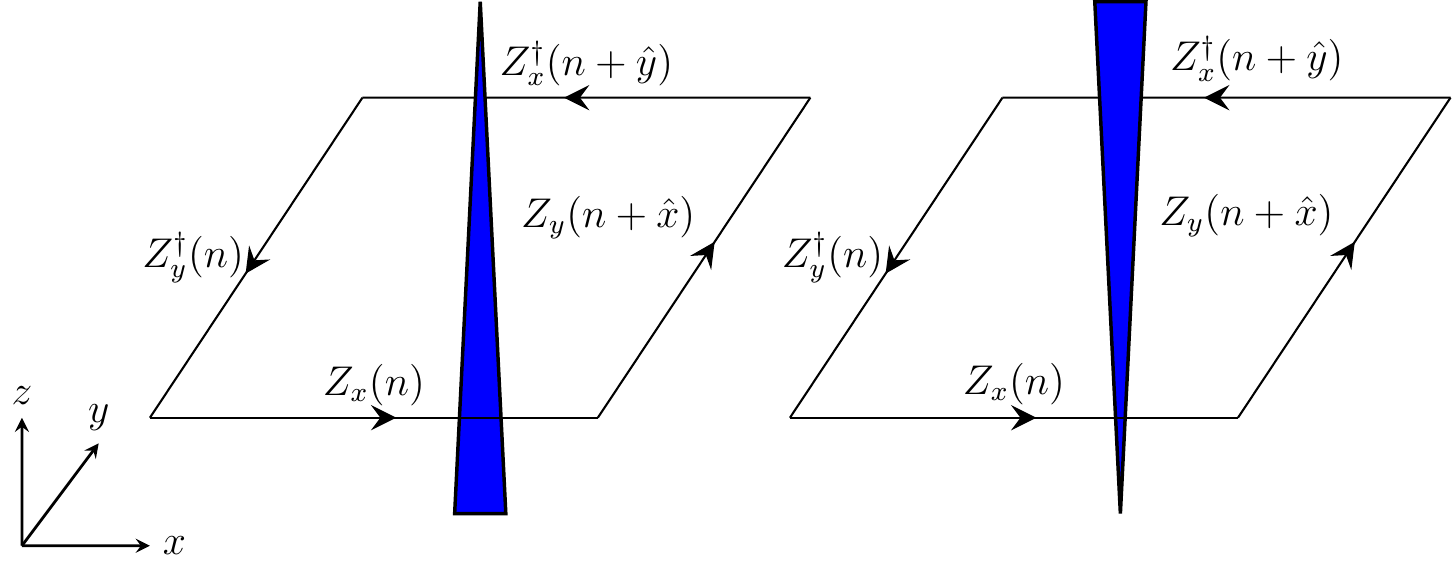}
  \caption{The spacial vortex plotting convention with $\lambda=4$. An $m=+1$
    vortex ({\bf left}) identified by the plaquette $P_{xy}(n)$ is rendered in
    the $\hat{z}$ direction. An $m=-1$ vortex ({\bf right}) identified by the
    same plaquette is rendered in the $-\hat{z}$ direction.}
  \label{fig:PlotConvention}
\end{figure}

To produce a 3D visualisation, one fixes the value of $\lambda$ in Eq.\eqref{eq:VortexPlaq} to be the dimension upon which slices are taken. The remaining three dimensions comprise the slice, such that the plaquettes now may be written as
\begin{equation}
  \label{eq:VortexPlaq3D}
  P_{ij}(\textbf{x})=\exp \left( \frac{2\pi i}{3}\,\epsilon_{ijk}\,m_{k\lambda}(\bar{\textbf{x}}) \right)\,,
\end{equation}
where the Latin indices enumerate the three dimensions orthogonal to the fixed
$\lambda$. Using this definition, a vortex is rendered as a jet of length $a$,
pointing in the $m_{k\lambda}(\bar{\vb{x}})\,\hat{k}$ direction that pierces the
$P_{ij}(x)$ plaquette. For example, if we choose $\lambda=4$, a $m=+1$ vortex
identified by $P_{xy}(n)$ would be rendered in the $+\hat{z}$ direction. This
rendering convention is illustrated in Fig.~\ref{fig:PlotConvention}.

A notable feature of $SU(3)$ centre vortices is the presence of vortex
branching. Due to the periodicity of the non-trivial centre phases in
$\mathbb{Z}_{3}$, one unit of positive centre charge is equivalent to two units
of negative centre charge. Hence, within a 3D slice a vortex line carrying
$m=+1$ charge may branch into two $m=-1$ vortex lines. Note that this process is
indistinguishable from three $m=+1$ vortex lines converging to the vacuum, as
illustrated in Fig.~\ref{fig:BranchingPointModel}. Recall that our visualisations
illustrate the directed flow of $m = +1$ charge. This is why these branching
points are also sometimes referred to as vortex monopoles and anti-monopoles in
the literature~\cite{Spengler:2018dxt}. This ambiguity in charge assignment has
important ramifications for centre vortex topology, as discussed in
Ref.~\cite{Engelhardt:1999xw}.

For the purposes of this work, we will refer to intersections of three or five
vortices as branching points. Intersections of four vortices occur at the
intersection of vortex lines and do not constitute vortex branching. They are thus
excluded from the branching point analysis. Finally, intersections of six vortices
could arise from either vortex branching or the intersection of three vortex
lines. As these situations are indistinguishable, for this work we will consider
these points to be branching points. However, it must be noted that the
occurrence of six-way branching points is so infrequent that this choice has an
insignificant impact on branching point statistics.

A straightforward nomenclature for referring to branching
points~\cite{Spengler:2018dxt} is to define the branching genus
$n_{\rm cube}(x\, |\, \hat{\mu})$. Here, $\hat{\mu}$ denotes the direction along which the
lattice has been sliced and hence identifies the remaining three coordinates,
$\hat{\imath}, ~ \hat{\jmath}, ~ \hat{k}$, that describe the location within the 3D slice.
Within the selected slice, we define $\textbf{x}^{\prime}$ to denote the dual
lattice site,
$\textbf{x}^{\prime} = \textbf{x} + \frac{a}{2}(\hat{i}+\hat{j}+\hat{k})$.
$n_{\rm cube}(x\, |\, \hat{\mu})$ then counts the number of vortices piercing the elementary
cube around $\textbf{x}^{\prime}$. Thus, we have the following interpretation
for the possible values of $n_{\rm cube}(x\, |\, \hat{\mu})$:
\begin{equation}
  \label{eq:BPGenus}
  n_{\rm cube}(x\, |\, \hat{\mu}) =
  \begin{cases}
      0 & \text{No vortex} \\
      2 & \text{regular vortex line} \\
      3,5,6 & \text{branching point} \\
      4 & \text{touching point}
   \end{cases}
\end{equation}

\begin{figure}
  \centering
  \includegraphics[width=\linewidth]{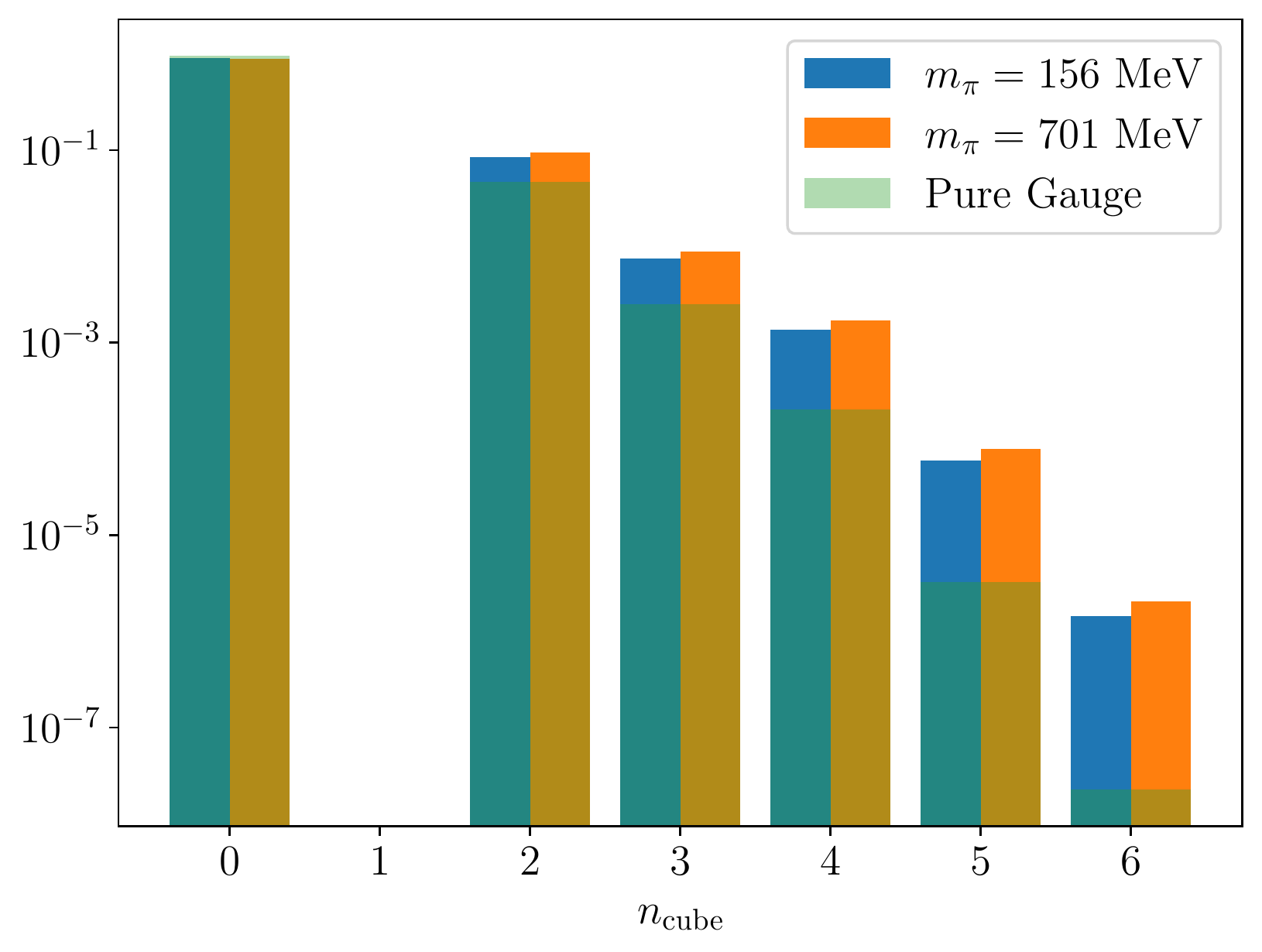}
  \caption{The distribution of branching point genera as defined in Eq.~\eqref{eq:BPGenus}.}
  \label{fig:BPGenusDist}
\end{figure}

The normalised distribution of values of $n_{\rm cube}$ across the three
ensembles is shown in Fig.~\ref{fig:BPGenusDist}. We observe that the
distribution of the higher genus values decreases monotonically for all
ensembles. The dynamical ensembles feature a greater probability of
high-multiplicity branching points. This predicts a greater vortex density for
these ensembles relative to the pure gauge case, as will be discussed in the
next section.

\section{Cluster Identification}\label{sec:LoopID}
It is well known that for $SU(2)$ gauge fields in the confining phase,
percolation of centre vortices can be used as an order parameter for the
transition from the confined phase to the deconfined
phase~\cite{Engelhardt:1999fd,Bertle:1999tw}. At a glance, the visualisations
constructed in Ref.~\cite{Biddle:2019gke} support this assessment, with a single
large connected vortex cluster clearly visible in each visualisation and only a
handful of separate smaller secondary clusters present. Studying the confinement
phase transition at the critical temperature will be the subject of future work.
However, it is of interest to build the necessary tools to perform such a study.
This requires us to quantitatively understand the degree to which a vortex
ensemble is dominated by a primary percolating cluster, as opposed to a
collection of smaller secondary clusters. To do this, it is necessary to develop
an algorithm that can trace these vortex lines and identify disconnected
clusters.

Such an analysis is quite straightforward in $SU(2)$, as $SU(2)$ vortices do not
permit branching points. This simplifies the algorithm, as each vortex cluster
consists of a single line that may be followed until it arrives back at its
starting location. In $SU(3)$, vortex branching demands that the algorithm track
multiple branching paths, and only terminates when there are no continuations
for every path. We describe such an algorithm here.

The starting point for the algorithm is to have all vortices in a 3D slice
stored along with their associated tip and base coordinates. With this
setup, the algorithm proceeds as follows:
\begin{enumerate}
\item \label{it:startalg} Choose an arbitrary vortex to start at. Mark it as visited and record it as belonging to an incomplete line segment.
\item \label{it:startloop} Considering the last vortex in each incomplete line segment, produce a list of all unvisited vortices touching this vortex (both base and tip, accounting for periodicity). Then mark them all as visited
\item Append one of the found vortices to the current segment. For all others, begin a new segment.
\item If there are incomplete segments, repeat from step~\ref{it:startloop} for each incomplete segment.
\item Once there are no unvisited touching vortices, mark the segment as complete.
\item If all segments are complete, the cluster is complete. Record all vortices in all segments as belonging to this cluster. Return to step~\ref{it:startalg}, selecting an unvisited vortex.
\item If there are no unvisited vortices, all clusters have been identified and the algorithm is complete.
\end{enumerate}
This algorithm can then be applied to each 3D slice to isolate all independent vortex clusters.

\begin{figure*}
  \centering
  \includegraphics[clip=true,trim=5.0cm 0.0cm 4.5cm 6.5cm,width=0.48\textwidth]{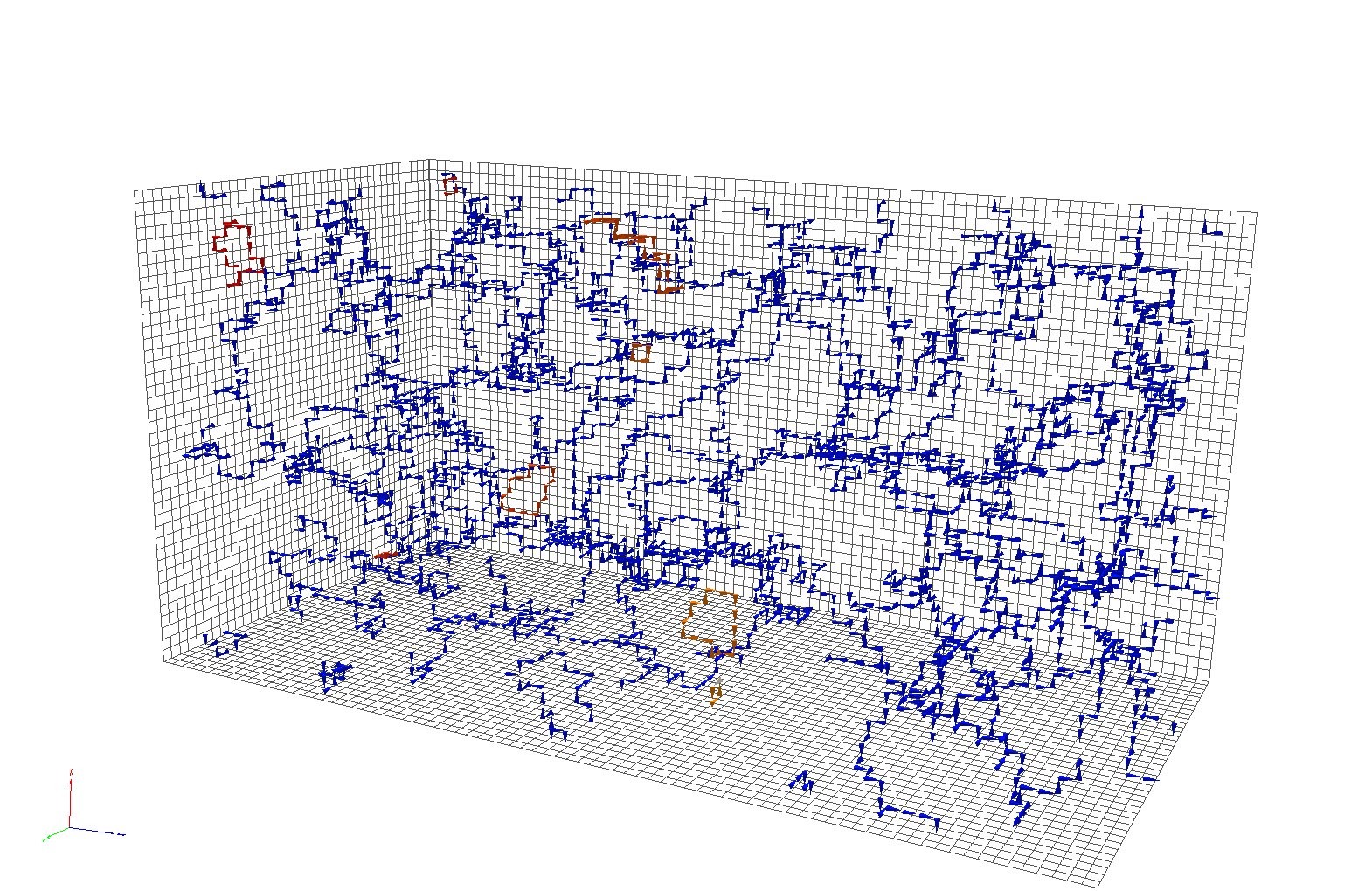} \quad
  \includegraphics[clip=true,trim=5.0cm 0.0cm 4.5cm 6.5cm,width=0.48\textwidth]{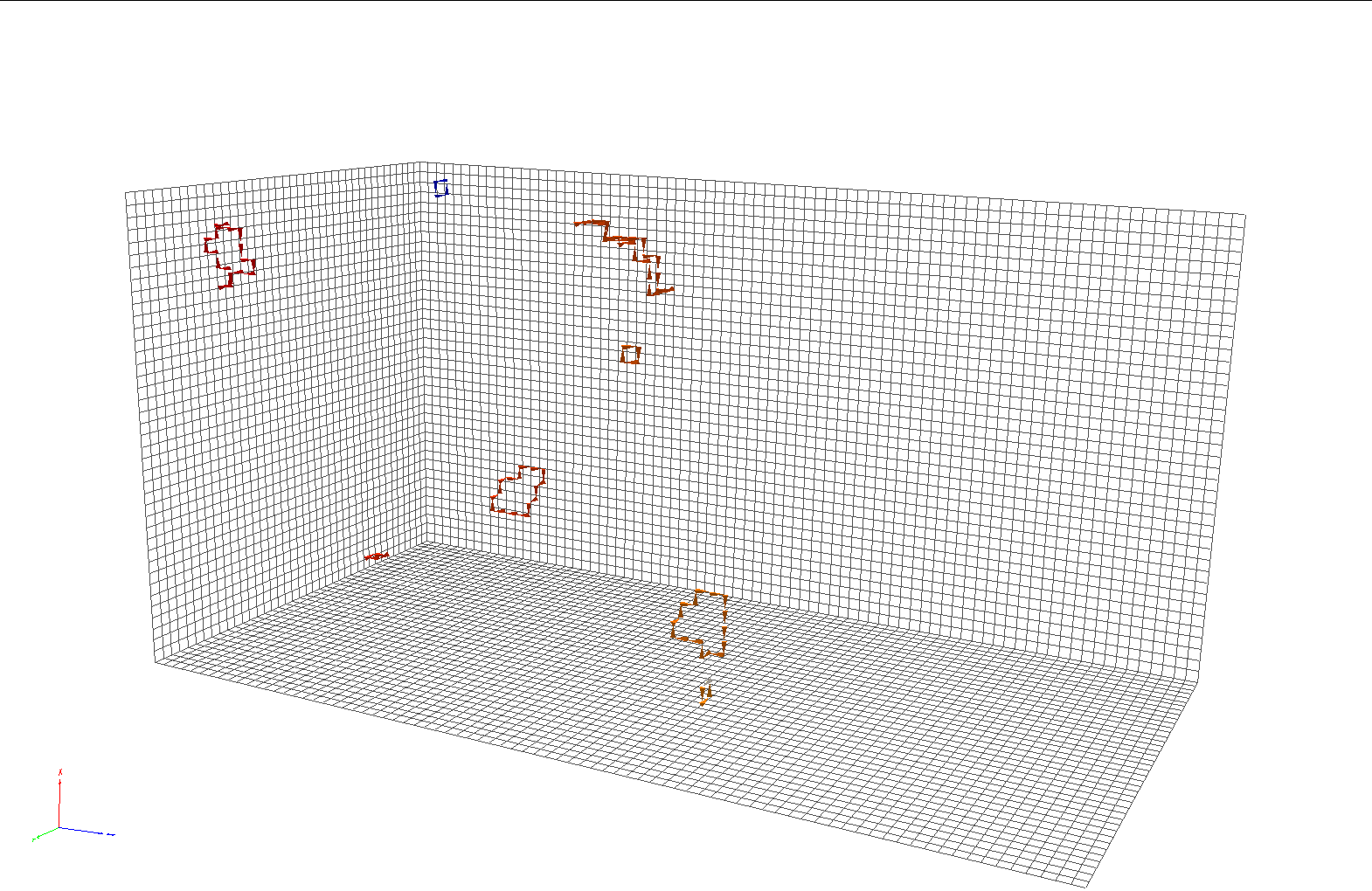} \\[6pt]
  \includegraphics[clip=true,trim=5.0cm 0.0cm 4.5cm 6.5cm,width=0.48\textwidth]{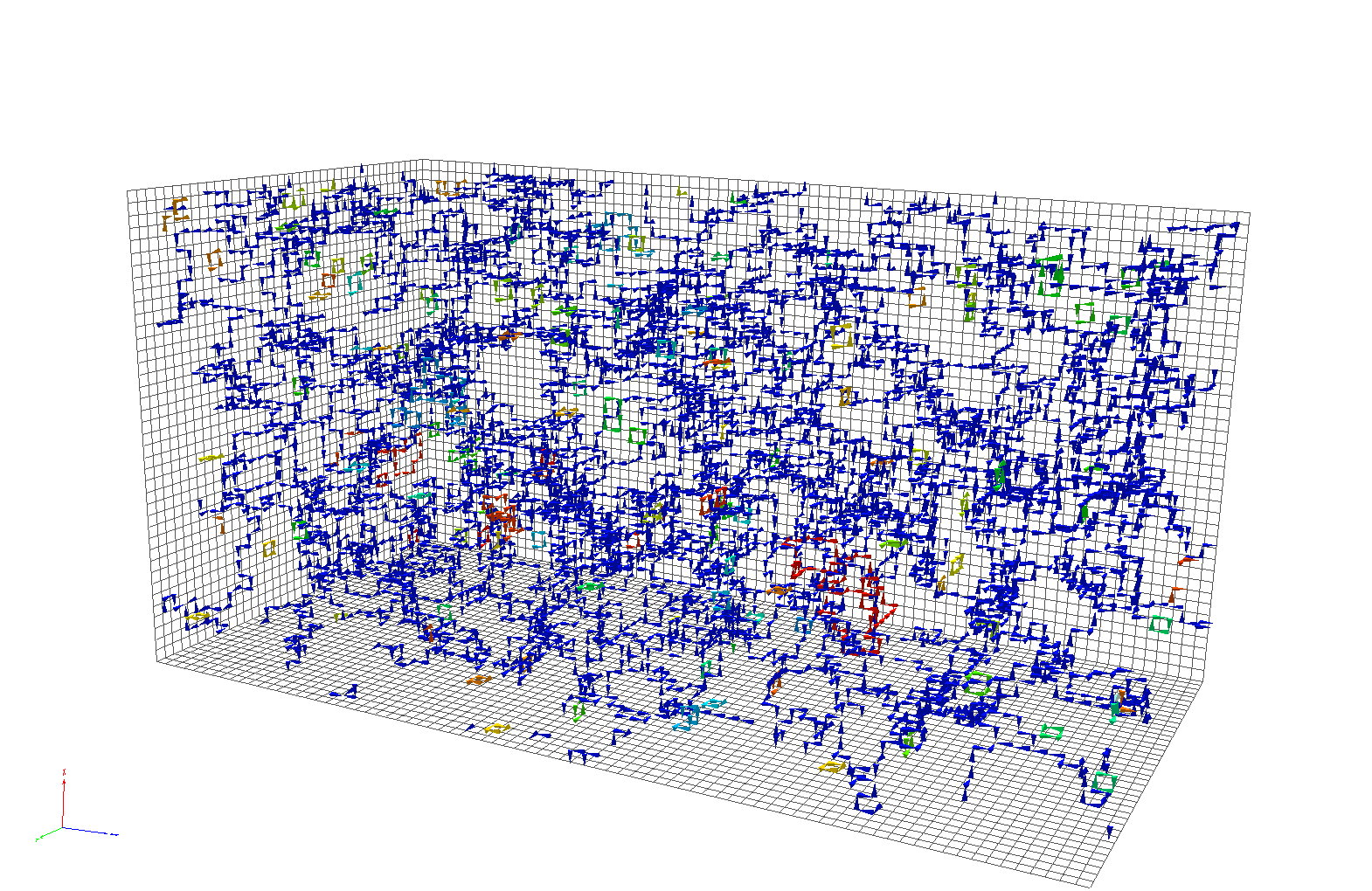} \quad
  \includegraphics[clip=true,trim=5.0cm 0.0cm 4.5cm 6.5cm,width=0.48\textwidth]{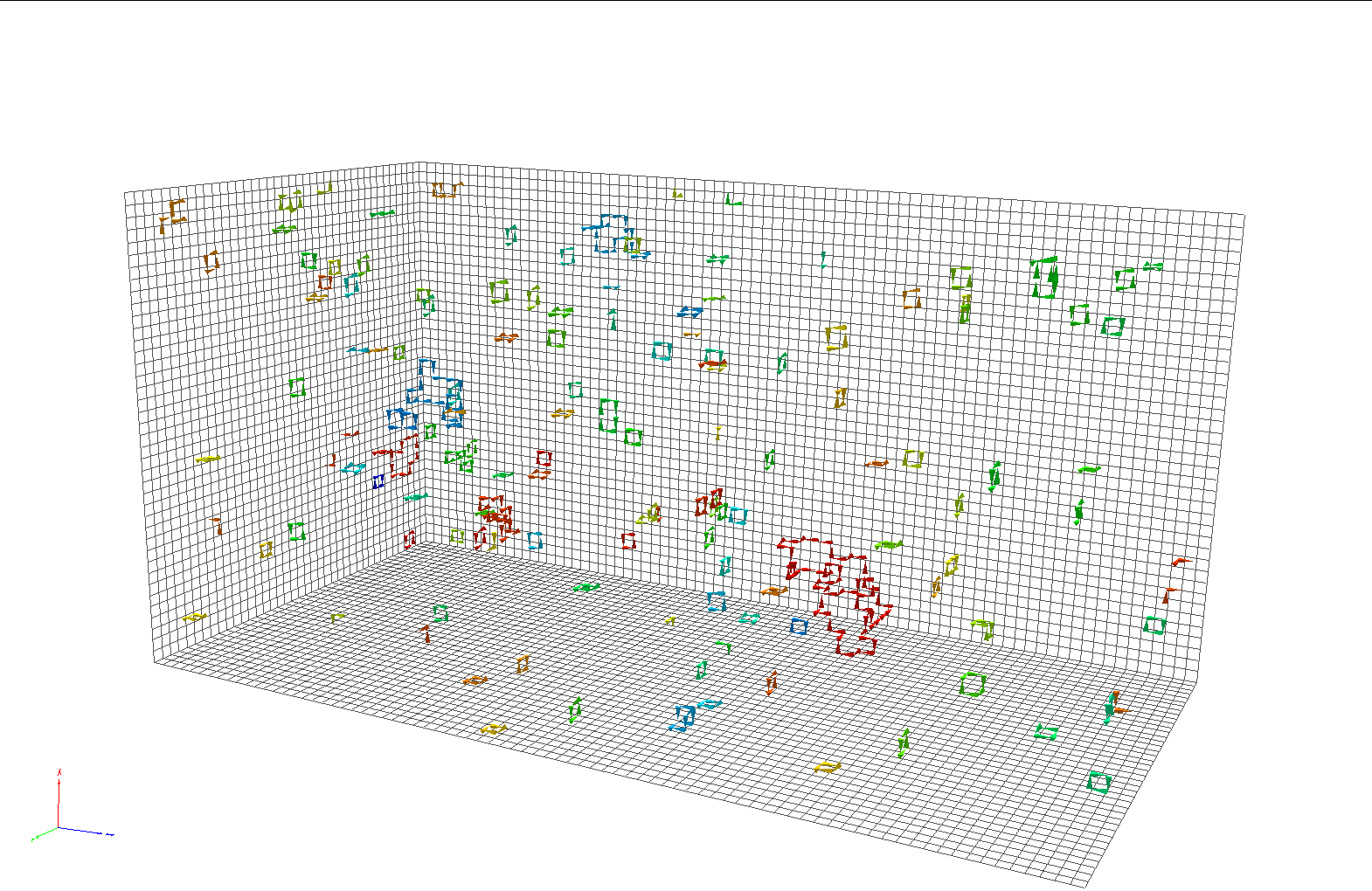}
  \caption{ (Top left) The centre vortex structure of a pure-gauge configuration.  
    (Top right) The pure-gauge vortex vacuum as shown in the top left panel with the primary percolating
    vortex cluster removed. 
    (Bottom left) The centre-vortex structure of a $2+1$ flavour dynamical-fermion configuration from the
    $m_{\pi}=156~\si{MeV}$ ensemble. 
    (Bottom Right) The dynamical vortex structure in the bottom-left panel with
    the primary percolating vortex cluster removed. Note the increased abundance of elementary
    vortex paths and the prevalence of branching points.
    In each panel, separate vortex clusters are rendered with different colours. These 3D models
    are generated with AVS scientific visualisation software~\cite{AVS:2020}.
    \textbf{Interactive} in the supplemental material.}
\label{fig:vis}
\end{figure*}

Employing this algorithm and our visualisation conventions defined in
Sec.~\ref{sec:Visualisations}, the pure-gauge vortex vacuum on a single slice
appears as in top-left panel of Fig.~\ref{fig:vis}. The interactive version of
this visualisation may be found in
Fig.~\ref{sup-fig:Primary-Secondary-PG-24.u3d}. As our investigation takes place
at zero temperature on a large volume lattice, the choice of slice direction
does not impact most intrinsic measurements, and as such we choose to present
plots obtained from slicing in the $\hat{x}$ direction. The only notable
exception is the size of the percolating cluster as it fills the 3D volume and
is therefore smaller for $\hat t$ slices. The choice of $\hat x$ will be assumed
for the remainder of this work unless stated otherwise. Numerical values
presented in tables will be averaged across all slice dimensions, where
applicable.

\begin{figure}
  \centering
  \includegraphics[width=\linewidth]{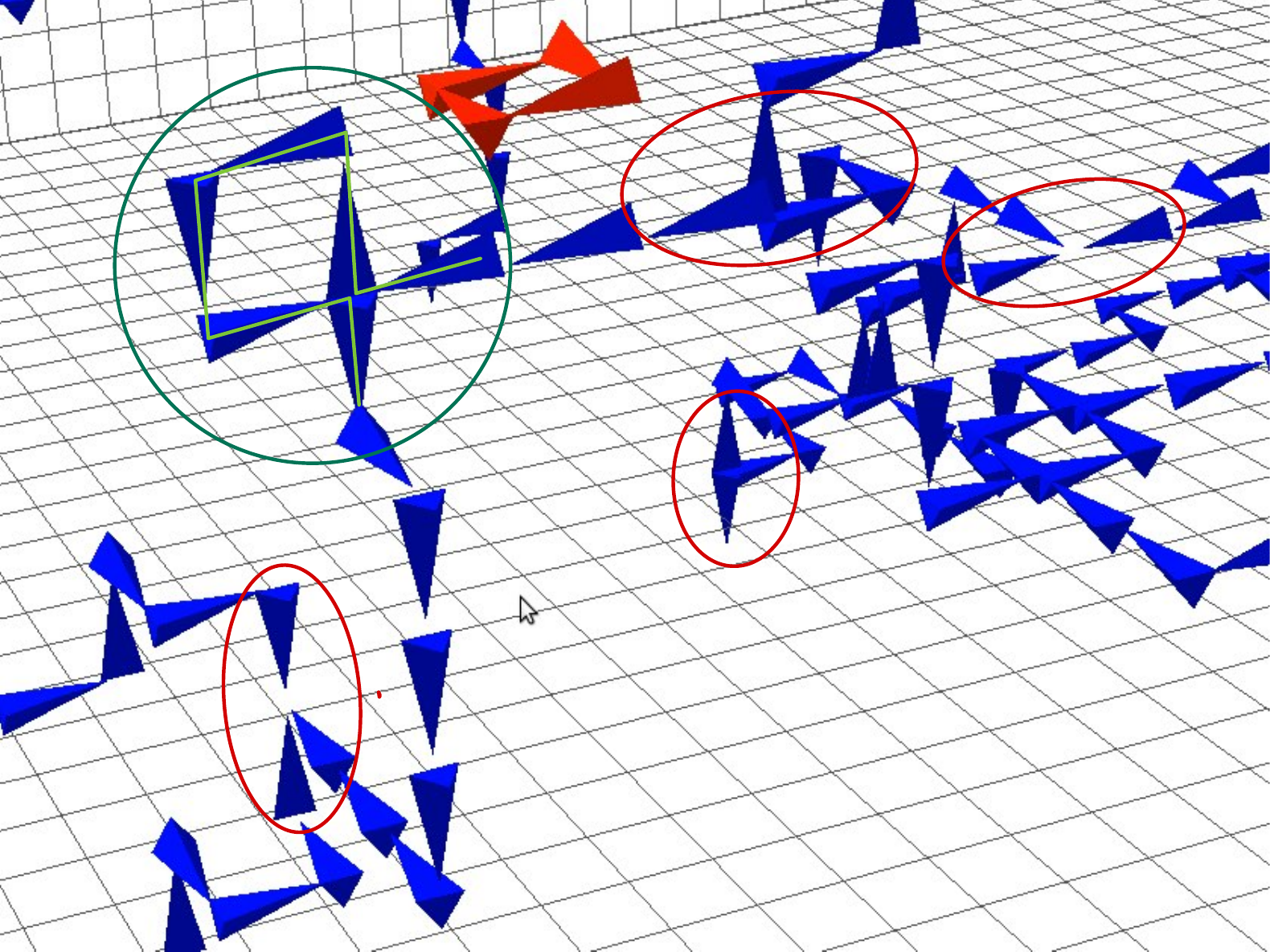}
  \caption{A collection of branching points (red ovals), a touching point (green circle) and a
    secondary loop (red jets) as they appear in our visualisations. Each jet illustrates the flow
    of $m = +1$ centre charge.}
  \label{fig:BranchingPointModel}
\end{figure}

We observe that indeed the vacuum is dominated by a single primary percolating cluster, with an
assortment of small secondary clusters also present. Branching points are readily observed within
the visualisation, as can be seen in Fig.~\ref{fig:BranchingPointModel} and in the interactive view
`Branching Points' in Fig.~\ref{sup-fig:Primary-Secondary-PG-24.u3d}.

The transition to full QCD leads to a marked shift in the behaviour of the
centre vortices, as can be seen from the vortex vacuum of the lightest pion mass
ensemble shown in the bottom-left panel of Fig.~\ref{fig:vis}. The interactive
version of this visualisation may be found in
Fig.~\ref{sup-fig:Primary-Secondary-DF-05.u3d}.  The total number of vortices
has increased significantly.

The dominance of a single vortex cluster is even more pronounced once it is
removed, as shown in the right-hand panels of Fig.~\ref{fig:vis} for the
pure-gauge (top) and dynamical-fermion (bottom) slices.  Almost all the vortex
matter is associated with the percolating cluster.  However, if we focus on
the dynamical-fermion secondary clusters in the bottom-right panel of
Fig.~\ref{fig:vis}, we see that the number of secondary clusters has increased
substantially when compared to the pure gauge ensemble.  Moreover, an increase
in the complexity of the secondary structures through branching-point clusters
is also evident.

These secondary clusters may also be explored in the interactive models given in
Figs.~\ref{sup-fig:Secondary-PG-24.u3d} and \ref{sup-fig:Secondary-DF-05.u3d}
for the pure-gauge and dynamical-fermion cases.  There several features are
highlighted in the ``Views'' menu and these views are also available in the full
vortex illustrations of Figs.~\ref{sup-fig:Primary-Secondary-PG-24.u3d} and
\ref{sup-fig:Primary-Secondary-DF-05.u3d}.

To gauge the relative sizes of the primary and secondary clusters,
we calculate the average total number of vortices per slice, $N_{\rm slice}$,
the average number of vortices associated with the primary cluster, $N_{\rm
  primary}$, and the average number of vortices associated with a secondary
cluster, $N_{\rm secondary}$.  $N_{\rm slice}$, $N_{\rm primary}$, and $N_{\rm
  secondary}$ for all three ensembles are presented in
Table~\ref{tab:vortexcount}. Note that the spatial values are obtained by
averaging across the three spatial dimensions acting as the slice
dimension. When $\hat{t}$ is selected for slicing the four dimensional volume,
the spatial volume is half that when a spatial direction is selected. As such,
the percolating cluster values in the $\hat{t}$ column are expected to be half
those in the spatial slicing column. 

Interestingly, we observe that $N_{\rm secondary}$ decreases in the presence of
dynamical fermions, indicating that the secondary clusters are smaller on
average. This is due to a proliferation of elementary plaquette vortex paths in
dynamical fermion QCD, as illustrated in the bottom-right panel of
Fig.~\ref{fig:vis}. 

We also see that $N_{\rm slice}$ and $N_{\rm primary}$ from the heavier
quark-mass ensemble are larger than the values calculated on the light
ensemble. This is likely a result of the fact that the heavier pion mass
configurations have a slightly larger physical volume. We can determine if this
is the case by considering the vortex density, $\rho_{\rm vortex}$.

\begin{table}[tb]
  \caption{The average number of vortices associated with: the total per 3D
    slice ($N_{\rm slice}$), the primary cluster ($N_{\rm primary}$), and a
    secondary cluster ($N_{\rm secondary}$), as calculated on the three
    ensembles. Separate averages are listed for the slicing dimension
    $\hat{\mu}$ being temporal or spatial.}
  \label{tab:vortexcount}
  \begin{ruledtabular}
    \begin{tabular}{lcc}
      \input{./include/loop_table.tex}
    \end{tabular}
  \end{ruledtabular}
\end{table}

The vortex density is calculated by considering the proportion of
plaquettes that are pierced by a vortex, $P_{\rm vortex}$. This is best calculated by first
defining an indicator function,
\begin{equation}
  v_{\mu\nu}(x) =
  \begin{cases}
      1, & P_{\mu\nu}(x) = \exp\left(\frac{\pm 2\, \pi\, i}{3}\right)\, I\\
      0, & P_{\mu\nu}(x) = I\, .
   \end{cases}
  \label{eq:vmunu}
\end{equation}
We then calculate the proportion of pierced plaquettes as,
\begin{equation}
  P_{\rm vortex} = \frac{1}{6\, V}\sum_{\substack{\mu,\, \nu\\ \mu < \nu}} \sum_{x}
  v_{\mu\nu}(x)\,,
  \label{eq:Pv}
\end{equation}
where the value $6$ counts the number of plaquettes associated with site $x$ in
four dimensions and $V = N_x\, N_y\, N_z\, N_t$ counts the number of sites in
the sum over $x$. The physical density is then given by,
\begin{equation}
  \label{eq:vortex-density}
  \rho_{\rm vortex}=\frac{P_{\rm vortex}}{a^{2}} \, .
\end{equation}

In the case where the vortex distribution is isotropic, the density derived in four dimensions is equal to the mean of the three-dimensional density when averaged over slices (such as in Fig.~\ref{fig:vis}).
We can decompose the lattice coordinates into a $1+3$-dimensional notation, $x = (w,\textbf{x}\, |\, \hat{\mu}),$ with $w$ corresponding to the index in the slicing dimension $\hat{\mu}$ and $\textbf{x}$ specifying the location within the corresponding hyperplane. 
Then the vortex density for slice $w$ along the dimension $\hat{\mu}$ is
\begin{equation}
  P_{3}(w,\hat{\mu}) = \frac{1}{3\, V_3(\hat{\mu})}\sum_{\substack{i,\, j \\ i < j,\, \neq \mu}} \sum_{\textbf{x}} v_{i j}(w,\textbf{x}\, |\, \hat{\mu})\,,
\end{equation}
where $v_{i j}(w,\textbf{x}\, |\, \hat{\mu})$ is the restriction of the indicator function in Eq.~\ref{eq:vmunu} to the relevant slice, $V_3(\hat{\mu})$ is the corresponding 3-volume (e.g. $V_3(\hat{x}) = N_yN_zN_t$), and the division by $3$ averages the number of plaquettes associated with each site in three dimensions.

Upon averaging over all $w$ slices in a given dimension and then averaging over the four slice
directions, one finds the following for the mean density
\begin{equation}
  \bar{P}_{3} = \frac{1}{3\, V}\frac{1}{4} \sum_\mu\sum_{\substack{i,\, j \\ i < j,\, \neq \mu}} \sum_{w,\textbf{x}} v_{i j}(w,\textbf{x}\, |\, \hat{\mu})\,,
\end{equation}
Noting that each plaquette has been counted twice in the sum over $i,\, j$ and
$\mu$, one recovers $P_{\rm vortex}$ of Eq.~(\ref{eq:Pv}).  Of course, in both cases,
the physical density is governed by the area of the plaquette as in
Eq.~(\ref{eq:vortex-density}).

The vortex densities from the three ensembles are shown in
Table.~\ref{tab:vortexdensity}. We see that the $\rho_{\rm vortex}$ is indeed larger on
the ensemble with the lightest pion mass, indicating a consistent trend of
increasing vortex density as the physical pion mass is approached from above.

\begin{table}[tb]
  \caption{The vortex density as calculated on the three ensembles. The
    proportion of pierced plaquettes, $P_{\rm vortex}$, the physical vortex
    density, $\rho_{\rm vortex}$, the proportion of branching points, $P_{\rm
      branch}$ and the physical branching point density, $\rho_{\rm branch}$ are
    presented.}
  \label{tab:vortexdensity}
  \begin{ruledtabular}
    \begin{tabular}{lcccl}
      \input{./include/vortex_density_table.tex}
    \end{tabular}
  \end{ruledtabular}
\end{table}

Another quantity of interest is the branching point density. This is obtained by
considering the fraction of elementary cubes within each 3D slice that contain a
branching point, $P_{\rm branch}$. Again, this is best calculated by first considering
the indicator function
\begin{equation}
  \label{eq:BPIndicator}
  b(x\, |\, \hat{\mu}) =
  \begin{cases}
      1, & n_{\rm cube}(x\, |\, \hat{\mu}) = 3, 5, 6\\
      0, & \text{otherwise}\, .
   \end{cases}
\end{equation}
The branching point proportion is then given by
\begin{equation}
  P_{\rm branch} = \frac{1}{4\, V}\sum_{\mu}\sum_{x}b(x\, |\, \hat{\mu})\, ,
\end{equation}
where $\mu$ sums over all four dimensions.
As this density is defined as an average over 3D cubes, the associated physical
density is 
\begin{equation}
\rho_{\rm branch}=\frac{P_{\rm branch}}{a^3} \, . 
\end{equation}
The branching point density is shown in Table~\ref{tab:vortexdensity}. Here we
observe that the branching point density follows the same trend as the vortex
density, namely that it increases with decreasing dynamical quark mass.

\begin{figure}
  \centering
  \includegraphics[width=\linewidth]{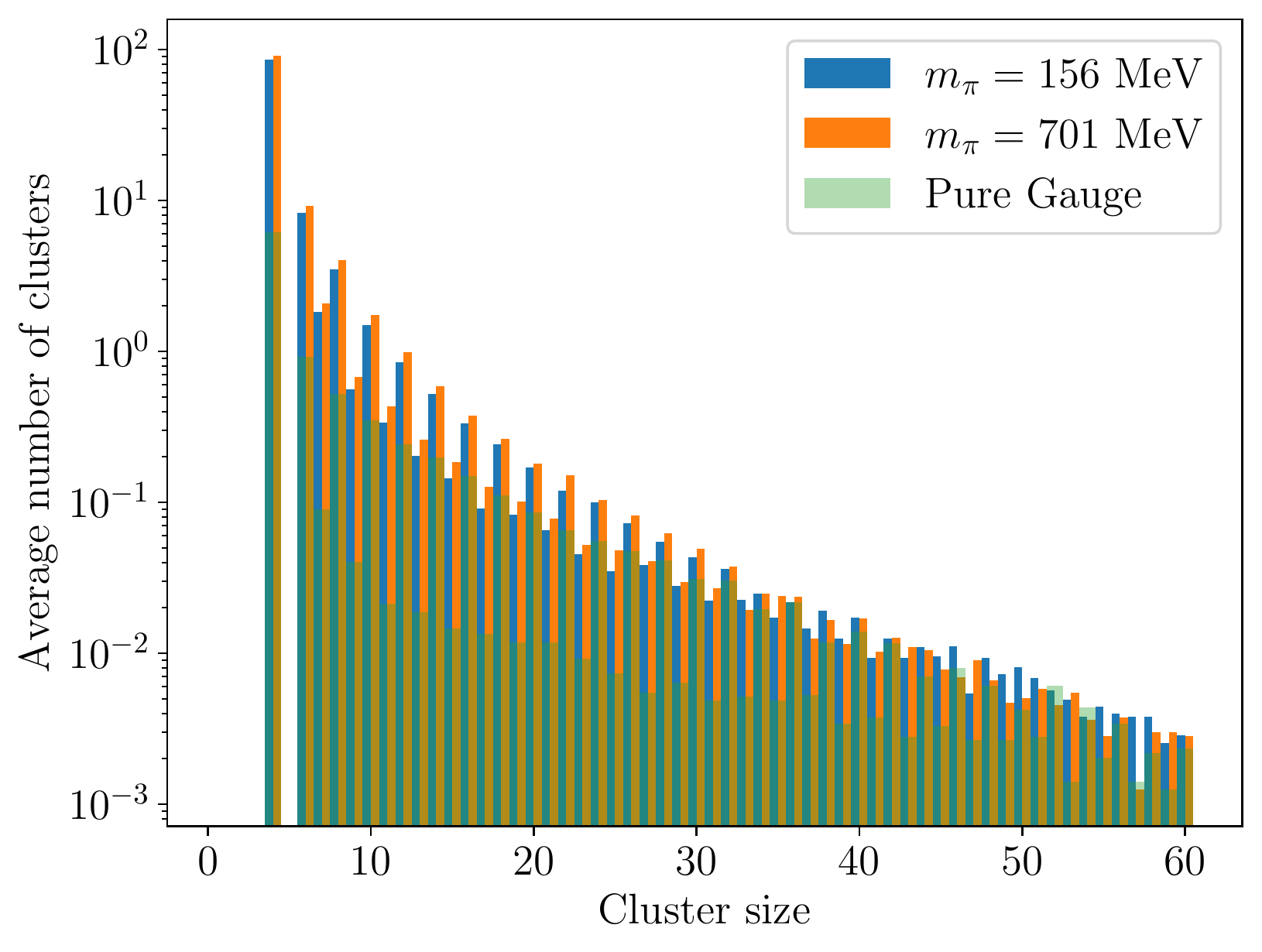}
  \caption{Average number of clusters of a given size per slice, up to a cutoff
    size of 60.}
  \label{fig:loophistnonorm}
\end{figure}

To quantify the change in the behaviour of $N_{\rm secondary}$ recorded in
Table~\ref{tab:vortexcount} we count the number of clusters of a given size and
average across slices and the ensemble. These results are shown in
Fig.~\ref{fig:loophistnonorm}. There are a number of interesting features
present here. Firstly, it is clear that it is not
possible to have clusters containing less than four vortices, and that it is
also not possible to have five vortices in a cluster. There is an interesting
trend that the number of clusters containing an even number of vortices is
higher than the number containing an odd number of vortices, especially at small
cluster sizes. This results in the alternating comb pattern present in
Fig.~\ref{fig:loophistnonorm}. This is a result of the fact that a branching
point is necessary for a cluster to contain an odd number of vortices. Hence,
this alternating pattern speaks to the presence of a `cost' associated with a
branching point, resulting in clusters containing branching points being less
probable than those without. This effect is mitigated as the cluster size
increases and the number of vortex arrangements leading to that cluster size
increases.

\begin{figure}
  \centering
  \includegraphics[width=\linewidth]{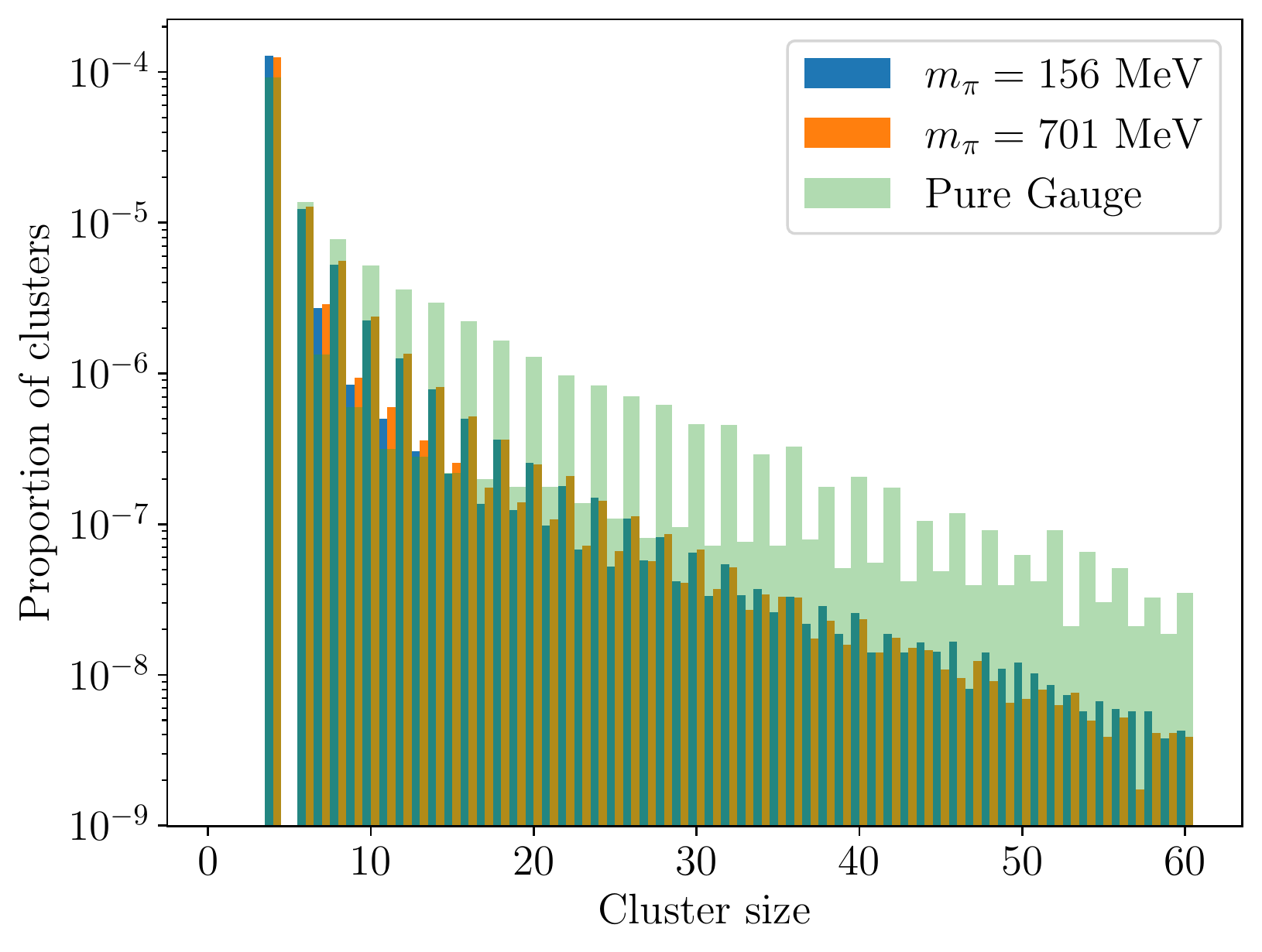}
  \caption{Proportion of clusters of a given size per slice, normalised by
    the total number of clusters in their respective ensemble.}
  \label{fig:loophistclusternorm}
\end{figure}

Comparing the different ensembles, we find that the number of clusters at each
size on the dynamical ensembles exceed almost all of the pure gauge clusters.
However, if we normalise the histogram by the total number of clusters found in
the ensemble, as shown in Fig.~\ref{fig:loophistclusternorm}, we find that the
pure gauge ensembles have a comparable or greater proportion of larger secondary
clusters present, perhaps due to the low vortex density. We observe that the
dynamical ensembles still retain a larger proportion of the smallest secondary
clusters.

\begin{figure}
  \centering
  \includegraphics[width=\linewidth]{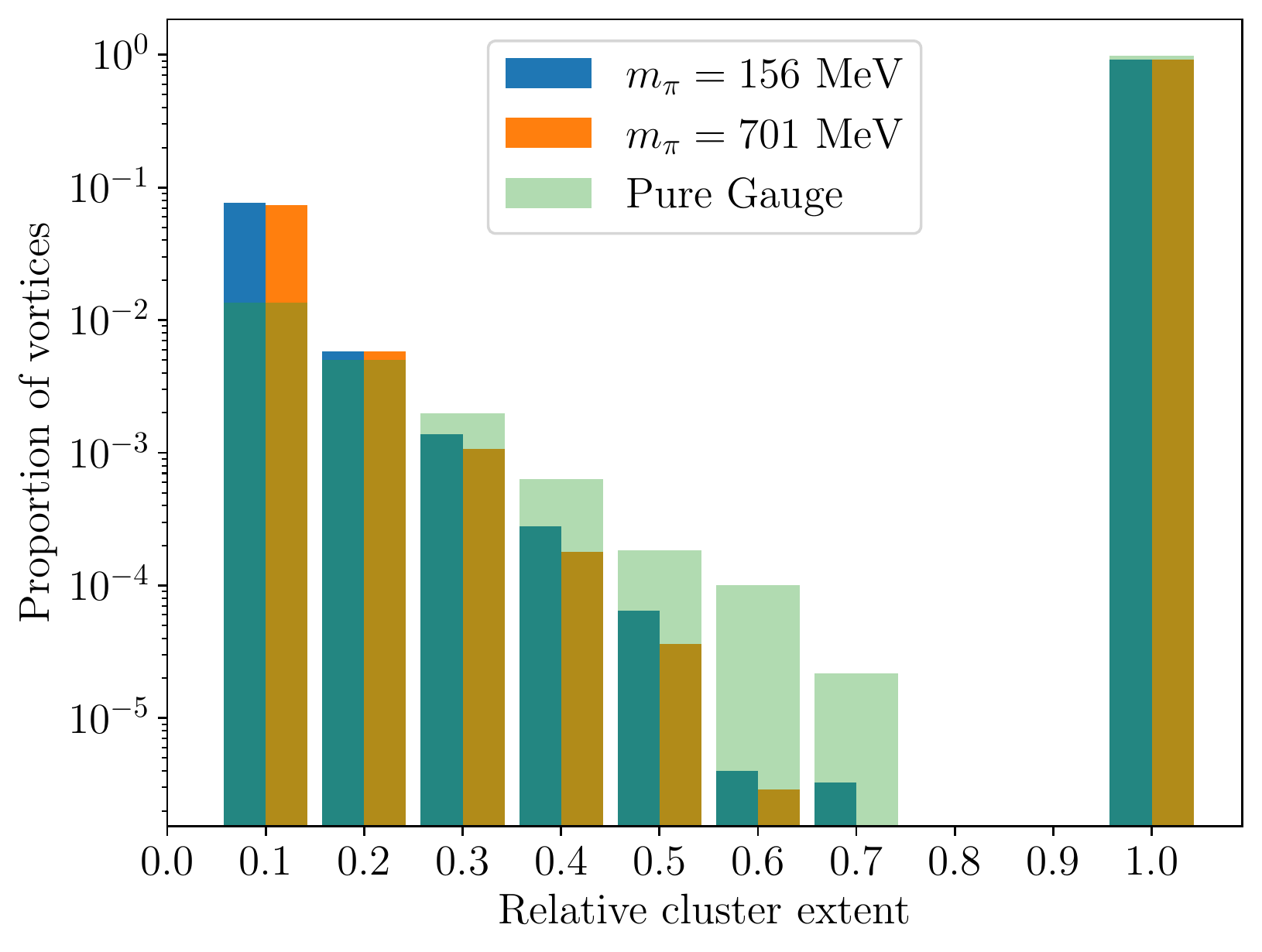}
  \caption{Histogram of the cluster extents relative to $L_{\rm max}$ for all
    three ensembles, as described in the text. It is clear that the vortex
    vacuum at zero temperature is dominated by a single percolating cluster, as
    can be seen by the dominance of the bin containing the clusters of maximal
    extent. Bin widths are $0.1$ and are centred at the tick marks of the
    $x$-axis. }
  \label{fig:clusterextent}
\end{figure}

We can measure the size of a cluster by defining the cluster extent as the
largest pairwise distance between vortices belonging to the same cluster, as
done in Ref.~\cite{Engelhardt:1999fd}. The cluster extents are binned, and the
content of each bin represents the average number of vortices in the associated cluster,
relative to the total number of vortices in the ensemble. The cluster extents
are normalised by the greatest distance on a $N_y\times N_z\times N_t$ slice of
a periodic lattice,
\begin{equation}
  \label{eq:largestextent}
  L_{\rm max} = \sqrt{(N_y/2)^2+(N_z/2)^2+(N_t/2)^2}\, .
\end{equation}
The results of this analysis for our
three ensembles is shown in Fig.~\ref{fig:clusterextent}.

The cluster extents shown in Fig.~\ref{fig:clusterextent} clearly demonstrate
that at zero temperature the $SU(3)$ vortex vacuum is dominated by a single percolating vortex
cluster, with only a minority of vortices comprising smaller secondary loops. It
is expected that this situation will change as the temperature exceeds the
critical temperature, as has been observed in $SU(2)$ gauge
theory~\cite{Engelhardt:1999fd}. We also observe that the pure gauge secondary
clusters tend to be larger than their dynamical counterparts.

We find that the vortex and branching point density significantly increases upon
the introduction of dynamical fermions. However, relative to the total number of
vortices present, the pure gauge sector contains a greater proportion of larger
secondary clusters than the dynamical case. Aside from the primary vortex
cluster, the dynamical vortex vacuum is dominated by an excess of very small
secondary clusters. The visualisations reveal significant branching-point
complexity in the large secondary clusters of the dynamical-fermion vortex
vacuum. Several features are highlighted in the ``Views'' menu of the
interactive figures provided in the supplemental material.

\section{Branching Point Graphs}\label{sec:BPGraph}
The cluster analysis presented in Sec.~\ref{sec:LoopID} enables us to gain
insight into the size of the primary and secondary vortex clusters. It is also
of interest to study the relationship between branching points, as these
structures are absent in $SU(2)$ where much of the analysis of vortex structure
has previously been performed. Furthermore, it is helpful to abstract the vortex
clusters such that we need not be concerned with their precise 3D coordinates.
To that end, we seek to represent vortex clusters as a directed graph, with
branching points acting as vertices and the edges being given by vortex lines,
with each edge weighted by the number of vortices in the line.

The algorithm to perform this graph construction starts with an identified
vortex cluster as defined in Sec.~\ref{sec:LoopID}. First, for each vortex we
evaluate whether it touches a point with $n_{\rm cube}(x\, |\, \hat{\mu}) \geq 3$ at
its tip, base, both or neither. Each branching or touching point should also
have a unique ID. The algorithm proceeds as follows:
\begin{enumerate}
  \item \label{it:startdist} Find an untraversed vortex with a branching/touching point
        at its base. If no untraversed vortex can be found, then we are done.
        Otherwise, set the found vortex to be the current vortex and mark it as
        traversed. Set the current inter-branching point distance to 1 and
        record the ID of the branching/touching point at the base.
  \item \label{it:vortexcheck} Check if the current vortex has a
        branching/touching point at its tip. If it does, create an edge between
        the saved branching/touching point ID and the ID of the
        branching/touching point at the tip with weight equal to the current
        inter-branching point distance. Return to step~\ref{it:startdist}.
  \item Otherwise, find the vortex with its base touching the tip of the current
        vortex and mark it as traversed. Set the new vortex to be the current
        vortex and add 1 to the inter-branching point distance. Return to
        step~\ref{it:vortexcheck}.
\end{enumerate}
The resulting graph encodes the separations between all branching and touching
points within a cluster without reference to the specific cluster geometry.

\begin{figure}
  \centering
  \includegraphics[height=0.8\linewidth]{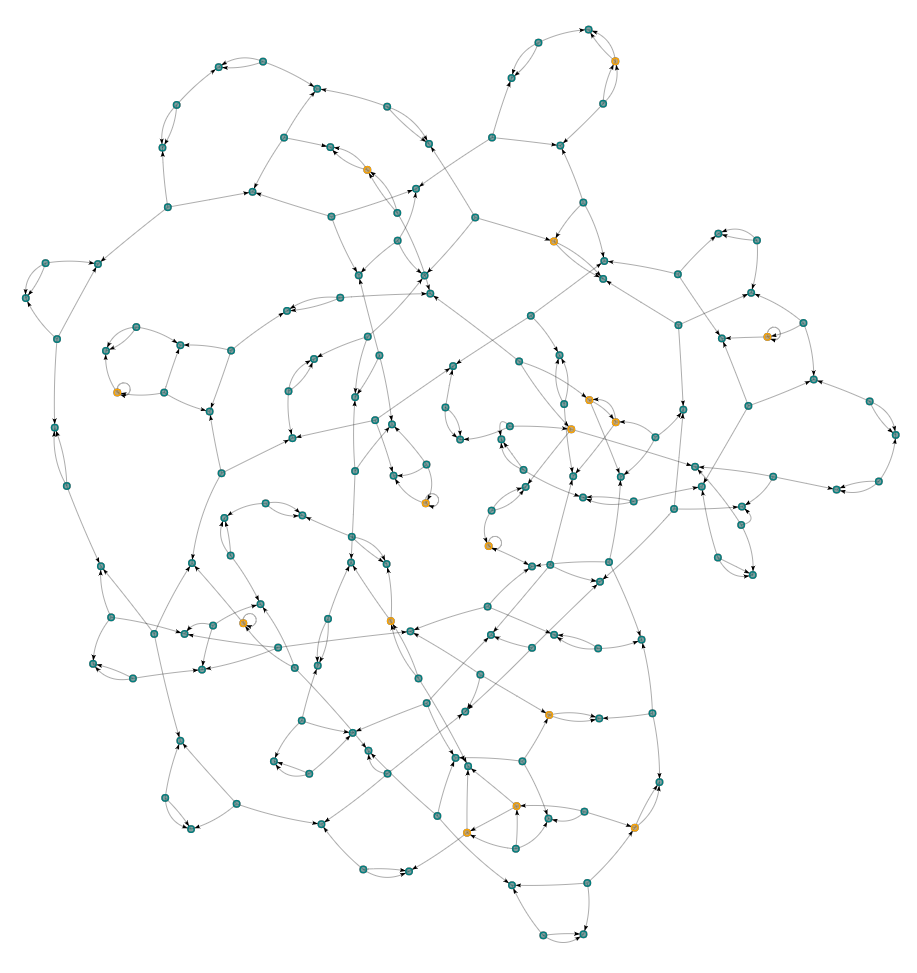}
  \caption{The pure-gauge primary vortex cluster from the slice shown in the top-left panel of
    Fig.~\ref{fig:vis} rendered as a graph. Branching/touching points are the vertices and
    connecting vortex lines are the edges. Blue vertices indicate three-way branching points and
    orange vertices indicate four-way touching points.  Visualisations were generated with the
    Pyvis visualisation package~\cite{pyvis:2018}.}
  \label{fig:PGgraph}
\end{figure}

\begin{figure}
  \centering
  \includegraphics[height=0.8\linewidth]{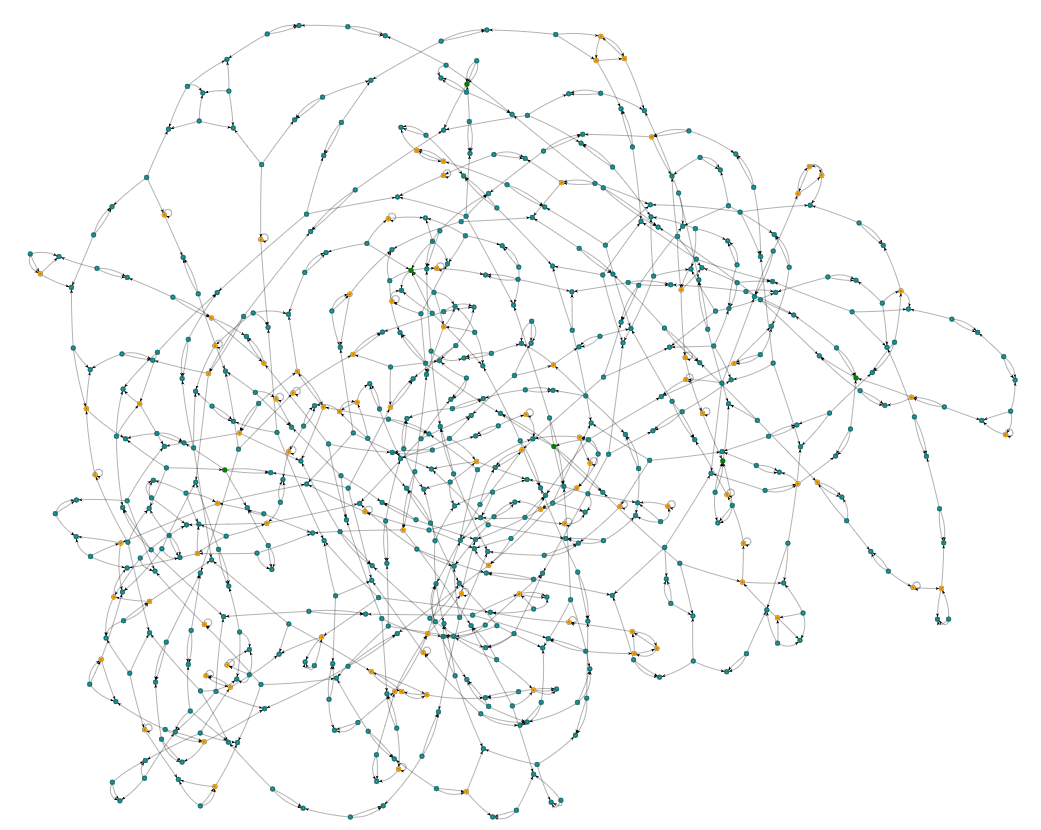}
  \caption{The $m_{\pi}=156 ~ \si{MeV}$ primary vortex cluster from the slice shown in the
    bottom-left panel of Fig.~\ref{fig:vis} rendered as a graph. Plotting conventions are as
    described in Fig.~\ref{fig:PGgraph}}
  \label{fig:dyngraph}
\end{figure}

Applying this algorithm to the primary clusters shown in Fig.~\ref{fig:vis} for
pure gauge and dynamical vacuum fields, we produce the graphs shown in
Figs.~\ref{fig:PGgraph} and \ref{fig:dyngraph} respectively. These
visualisations clearly demonstrate the significant increase in vortices and
branching points present on the dynamical configurations.

\begin{figure}[tb]
  \centering
  \includegraphics[width=0.95\linewidth]{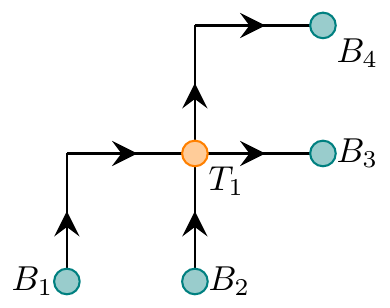}
  \caption{An example of how the touching point $T_1$ introduces ambiguity into
    the distance between branching points, $B_{i}$. $B_{1}$ can connect to
    either $B_{3}$ or $B_{4}$, with $B_{2}$ then connecting to $B_{4}$ or
    $B_{3}$ respectively. This would result in either distances of 4, 2 or 3, 3
    being recorded by our algorithm, depending on the order of traversal.}
  \label{fig:TouchAmbiguity}
\end{figure}

\begin{table*}
  \caption{The average distance between branching points, $d$, the same distance
    in physical units, $\Delta$, the average number of edges per graph, $n_{\rm
      edges}$, and the average number of edges per node, $n_{\rm edges}/n_{\rm
      nodes}$.}
  \label{tab:graphstats}
  \begin{ruledtabular}
    \begin{tabular}{lddcdd}
      \input{./include/graph_stats_table.tex}
    \end{tabular}
  \end{ruledtabular}
\end{table*}

Utilising this new construction, we wish to determine a measure of the
separation between connected branching points. A pair of branching points may be
connected via multiple vortex lines, and these lines may also pass through
touching points that we wish to exclude from the calculation. The presence of
these touching points makes it impossible to devise a unique distance between
two branching points, as this distance will depend on the manner in which the
touching point is traversed, as shown in Fig.~\ref{fig:TouchAmbiguity}. Instead,
we devise an algorithm for calculating the inter-branching point distance that
enables a random selection of directions with which to traverse these touching
point vertices. The algorithm proceeds as follows.
\begin{enumerate}
  \item \label{it:startgraph}Randomly choose a branching point vertex with
        untraversed outgoing edges. Record the vertex as the first
        in a path. Set the current path length to 0. If there is no vertex with
        an untraversed outgoing edge then we are done.
  \item \label{it:startrecur}Randomly choose an untraversed outgoing edge to
        follow to a new vertex. Mark the chosen edge as traversed, add the new
        vertex to the current path and add its length to the path length.
  \item If this edge arrives at a branching point, store the path and the
        current path length and return to step~\ref{it:startgraph}.
  \item If the edge arrives at a touching point, repeat from
        step~\ref{it:startrecur} with the new vertex as the starting vertex.
\end{enumerate}
The end result of this algorithm is a list of paths between branching points
that permit the ability to pass through touching points. However, not all edges
will be traversed by this method, as the presence of touching points allows for
cycles to emerge from these paths. Fortunately, due to conservation of vortex
flux, any cycle emerging from a given path will return to that same path. Hence
to rectify the algorithm, we simply need to traverse all cycles on a given path
and add their length to the existing length. This is done by performing a
modified depth-first search on each vertex to traverse any cycles that were
omitted from the above method. Pseudocode for this search on a single vertex is as
follows:

\begin{minipage}{\linewidth}
\begin{lstlisting}[language=python]
function dfs(this_vertex, path):
  for edge in this_vertex.edges:
    if (edge is not traversed
        and edge is outgoing):
      path.length += edge.length
      edge.traversed = True
      next_vertex = edge.end
      if next_vertex is not this_vertex:
        dfs(next_vertex, path)
\end{lstlisting}
\end{minipage}

The path lengths now accurately represent the distance between branching points.
This concludes our determination of the branching point separations. Note that
because of the inherent ambiguities in the branching point graphs, the solution
is not unique. We determine whether the impact of this randomness is significant
in the ensemble average choosing a single calculation of the distances as a
reference, then repeating the distance calculation nine further times with
different random seeds. We then use the Kolmogorov-Smirnov
test~\cite{hodges1958significance} to determine the equality of the different
distributions. We find that the test statistic for all ensembles is of order
$10^{-5}$, with corresponding $p$-values consistent with 1. Thus we are
satisfied that the variance in this distance measure is negligible in the ensemble
average, and we are therefore justified in considering it a useful measure of
branching point separation.

The average separation, $d$, for each ensemble is presented in
Table~\ref{tab:graphstats}. The physical separation $\Delta = a\, d$ is also
determined. Here we see that there is a consistent trend of decreasing average
separation with decreasing pion mass. This coincides with our determination of
the branching point and vortex densities, as a higher density suggests a smaller
separation between points.

We also present the average number of edges in the graphs, $n_{\rm edges}$, and
the average number of edges per node, $n_{\rm edges}/n_{\rm nodes}$ in
Table~\ref{tab:graphstats} as measures of the complexity and structure of the
graphs. We observe that, as expected, the number of edges substantially
increases upon the introduction of dynamical fermions. The number of edges per
node is close to $1.5$ for all ensembles, as the majority of edges emerge from a
three-way branching point and terminate at another three-way branching point.
However, the number of edges per node is larger on the dynamical ensembles,
likely due to the increase in vortex density resulting in a higher number of
vortex intersections.

\begin{figure*}
  \centering
\begin{tabular}{cc}
  \subfloat[Pure gauge.]{
    \includegraphics[width=0.43\textwidth]{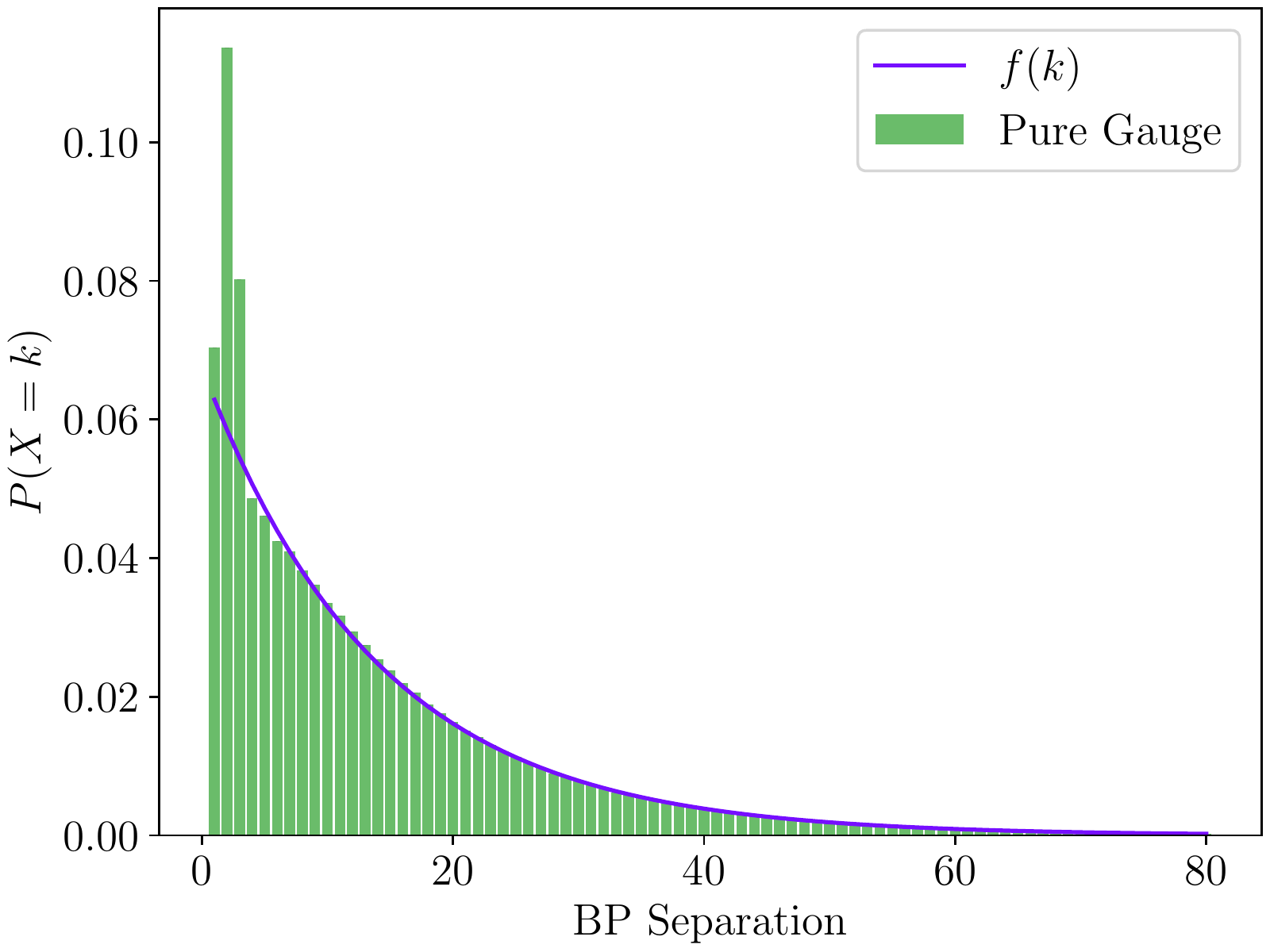}
  }  &
  \subfloat[Pure gauge, log scale.]{
    \includegraphics[width=0.43\textwidth]{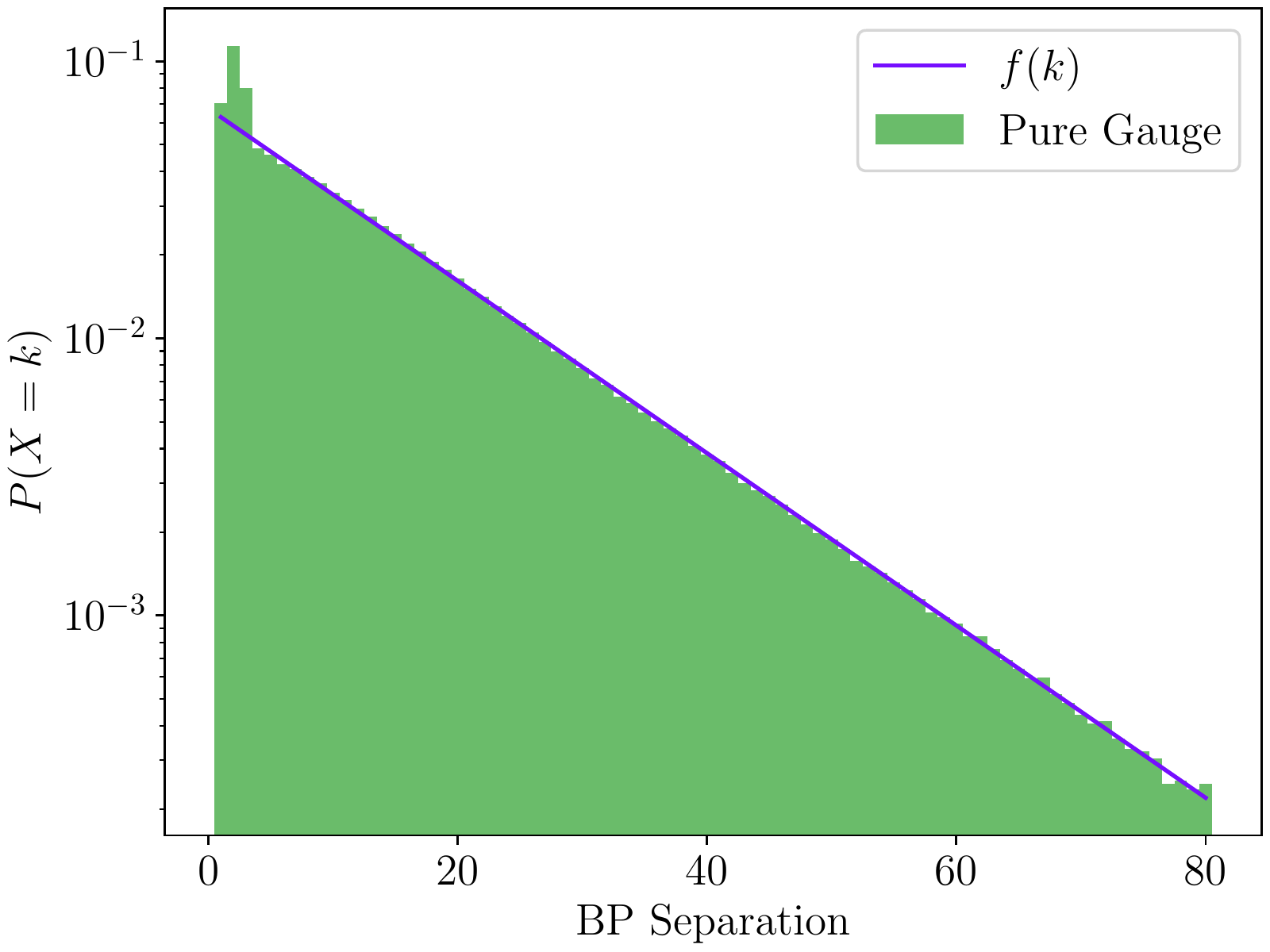}
  }\\

  \subfloat[$m_{\pi}=701~\si{MeV}$.]{
    \includegraphics[width=0.43\textwidth]{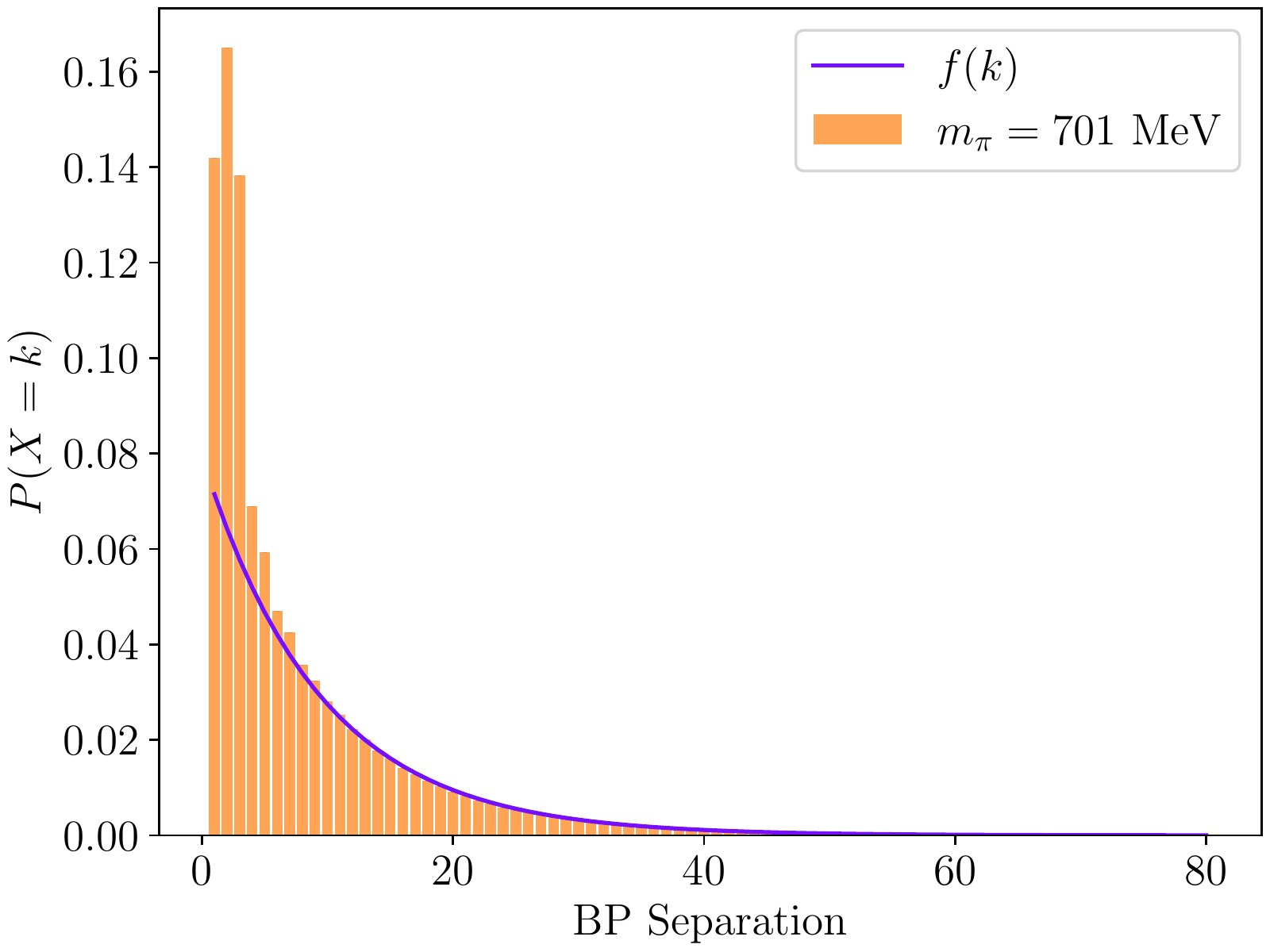}
  }  &
  \subfloat[$m_{\pi}=701~\si{MeV}$, log scale.]{
    \includegraphics[width=0.43\textwidth]{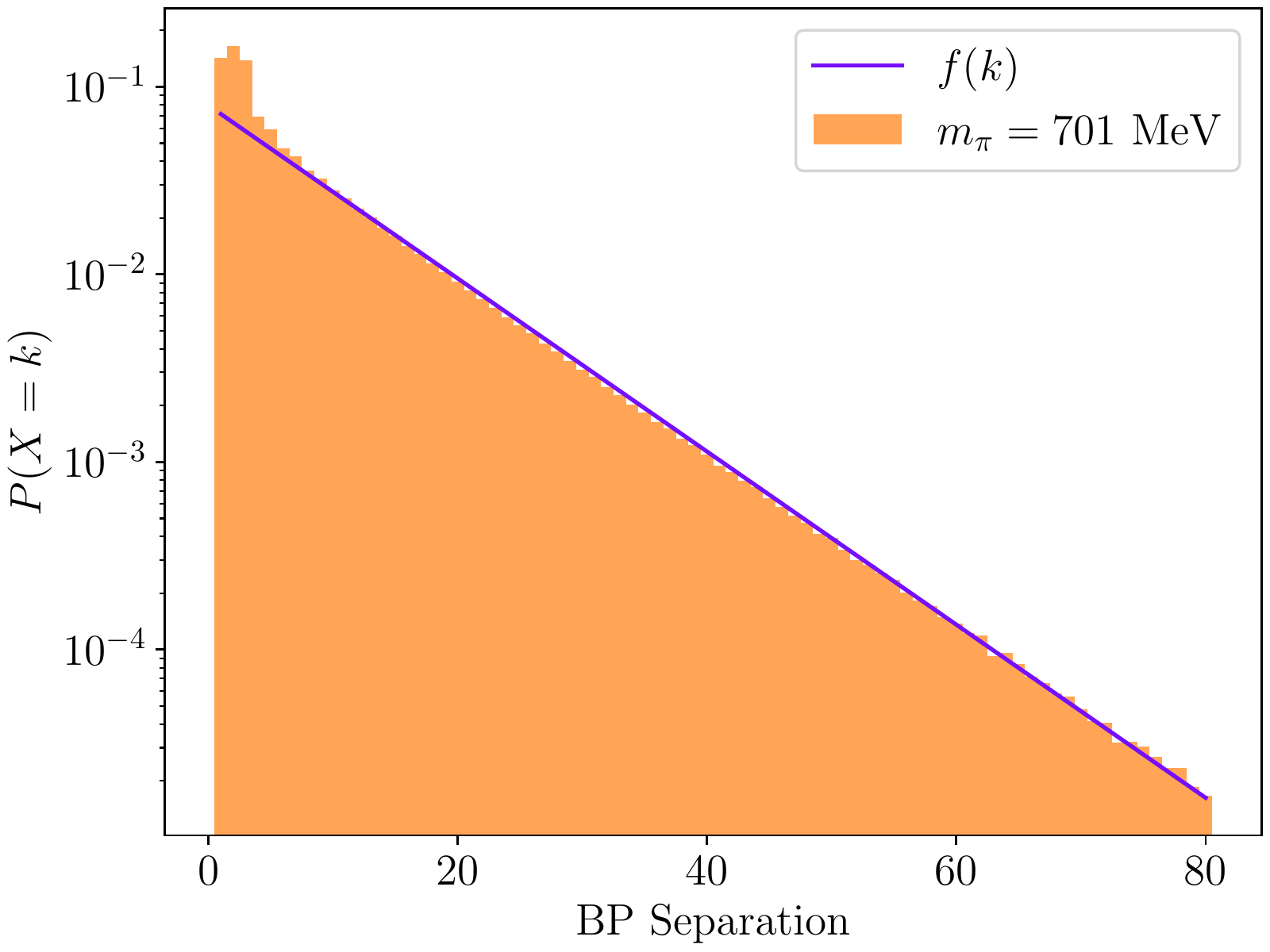}
  }\\

  \subfloat[$m_{\pi}=156~\si{MeV}$.]{
    \includegraphics[width=0.43\textwidth]{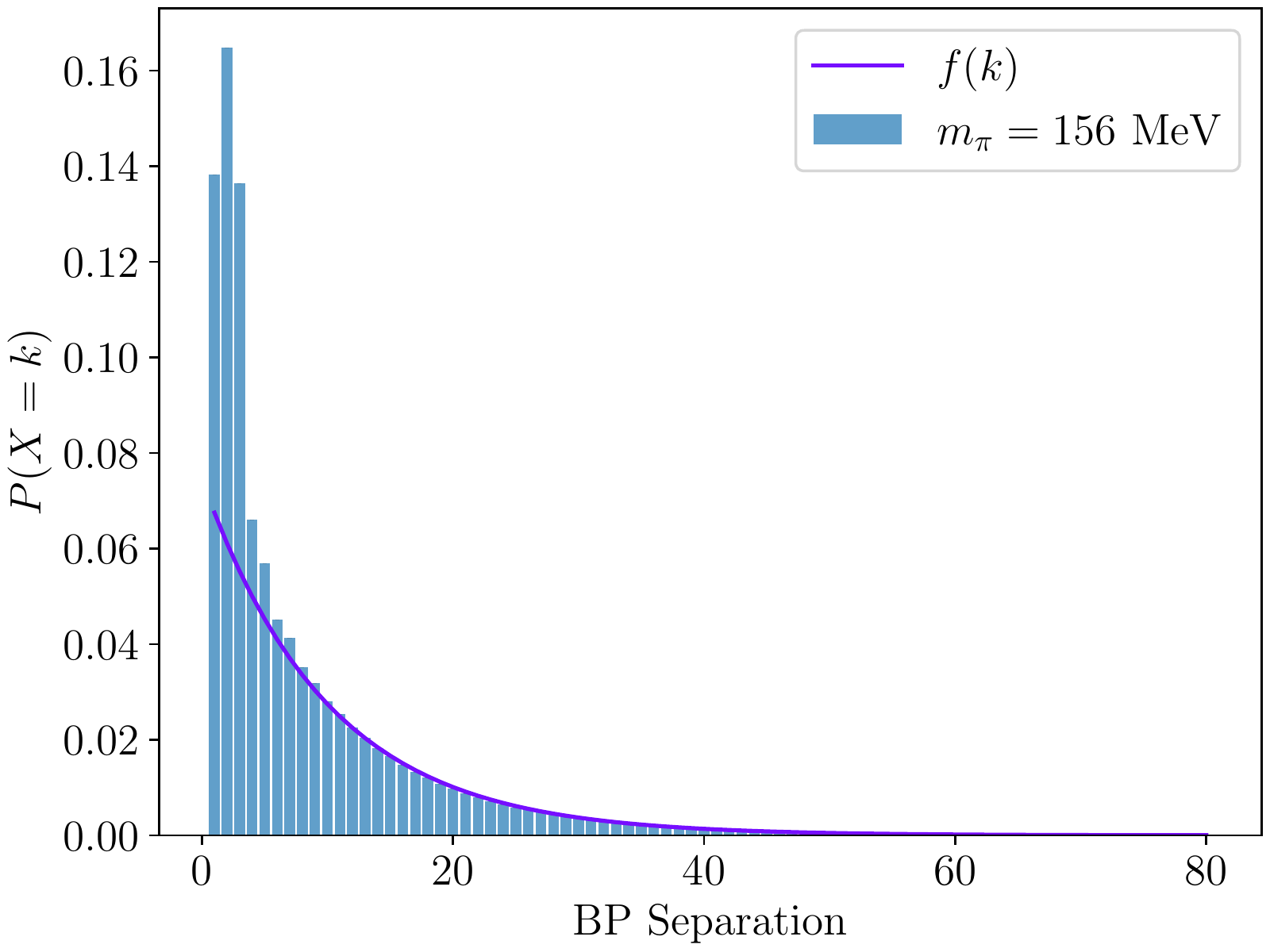}
  }  &
  \subfloat[$m_{\pi}=156~\si{MeV}$, log scale.]{
    \includegraphics[width=0.43\textwidth]{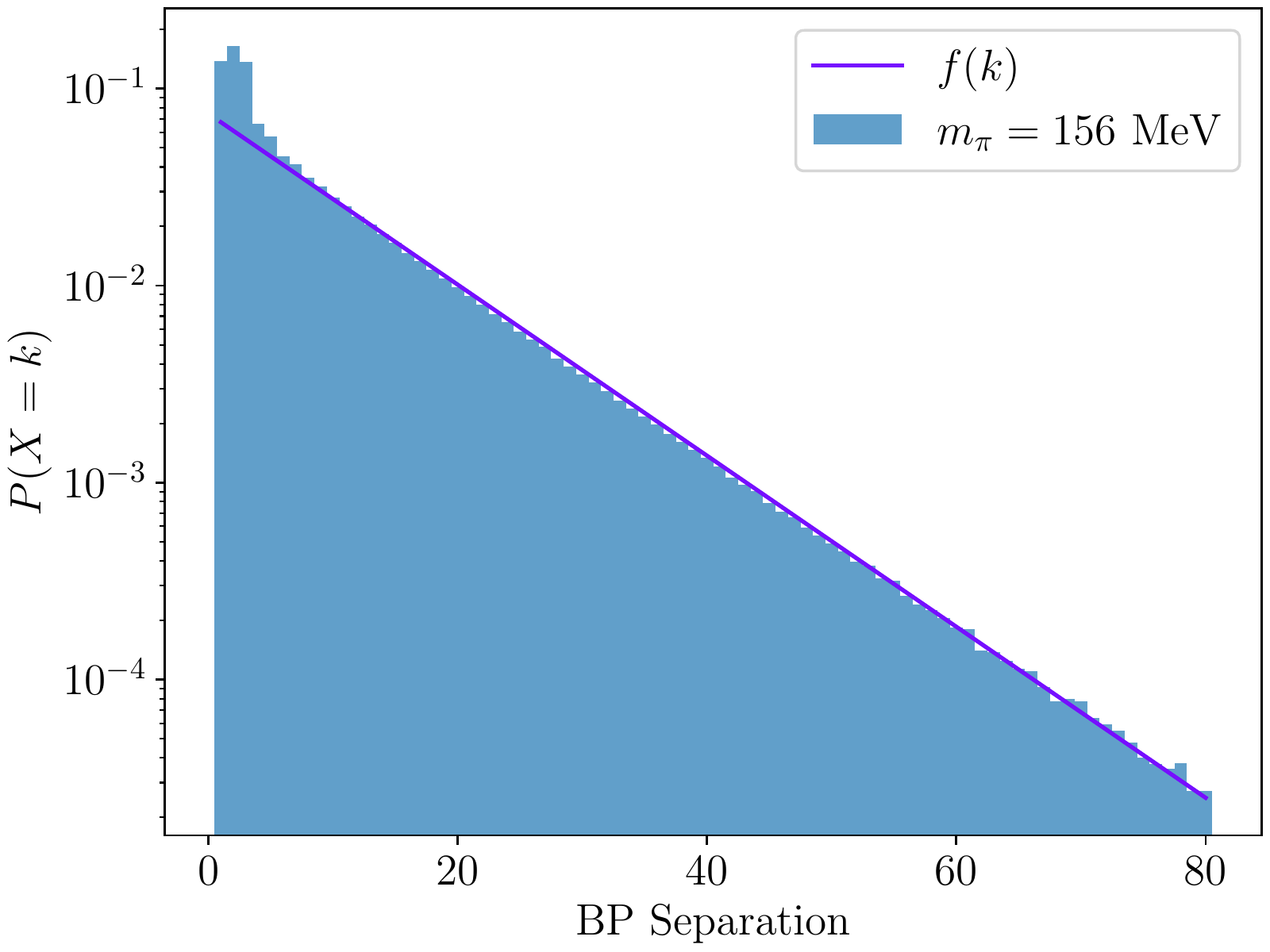}
  }\\
\end{tabular}
  \caption{ Normalised branching point (BP) separations from all
    ensembles, along with the corresponding fit to $f(k)$ given in
    Eq.~\eqref{eq:LinearFunc}.} 
  \label{fig:DistHistAll}
\end{figure*}

The distribution of branching point separations is shown in
Fig.~\ref{fig:DistHistAll}. The results are normalised by the total number of
vortex paths considered, such that the histogram has unit area. Apart from an
enhancement of the smallest branching point separations, the distances are
exponentially distributed. This distribution is consistent with a constant
branching probability, {\it i.e.} the probability of branching at the next link
of a vortex chain is independent of the length of the vortex chain.

This supports a previous conjecture for the interpretation of vortex
branching~\cite{Langfeld:2003ev,Spengler:2018dxt}: that a vortex can be
considered to have some fixed rate of branching as it propagates through
space-time. This interpretation allows for vortex branching on the lattice to be
considered as a binomial random variable $X$ with some probability of branching,
$q$. Thus, the probability of branching after $k$ lattice plaquettes is given by
the geometric distribution
\begin{equation}
  \label{eq:rate}
P_{k} = q\,(1-q)^{k-1}\,.
\end{equation}

Typically, one estimates the rate of a binomial random variable by evaluating
$q = 1/\bar{X}$, where $\bar{X} = \sum_{k}k\, P_{k}$. However, due to the
deviations from linearity found at small separations in the log-distributions
shown in Fig.~\ref{fig:DistHistAll}, this measure fails to capture the true rate
of branching. To account for this, we instead fit a linear function,
\begin{equation}
  \label{eq:LinearFunc}
  f(k) = \alpha -\beta\, k\,,
\end{equation}
to the log of the distribution of branching point separations for $k > 3$. 
The result of this fit for each ensemble is plotted in
Fig.~\ref{fig:DistHistAll}.

Of course, for a normalised distribution, $\alpha$ is constrained by $\beta$.
However, the significant non-exponential behaviour for $k \leq 3$ spoils the
exponential normalisation constraint.  Thus $\alpha$ is introduced to
accommodate for this, and we refer to $\beta$ describing the $k$ dependence to
determine the branching probability $q$.

The parameters of this fit are related to the $\log$ of the binomial rate 
\begin{equation}
\log(P_k) =  \log(q) - \log(1 - q) + \log(1 - q)\, k = \alpha - \beta\, k\, .
\end{equation}
Equating the coefficients of the terms linear in $k$, we resolve the branching rate
\begin{equation}
  \label{eq:7}
  q = 1 - e^{-\beta}\, .
\end{equation}
Note, for small $\beta$, $q = \beta$.  This rate can be converted to a physical
quantity by then considering the rate per unit length, $\lambda=q/a$. All fitted
parameters are calculated on 200 bootstrap ensembles, with errors determined via
the bootstrap variance.

The rate described above can then be compared to the naive rate,
$q_{\rm naive}$, calculated by considering the number of cubes
containing branching points divided by the number of cubes pierced by two or
more vortices. Defining
\begin{equation}
  c(x\, |\, \hat{\mu}) =
  \begin{cases}
      1, & n_{\rm cube}(x\, |\, \hat{\mu}) \neq 0\\
      0, & \text{otherwise}\,,
   \end{cases}
\end{equation}
and recalling the branching point indicator defined in Eq.~\eqref{eq:BPIndicator}, we define the naive rate to be,
\begin{equation}
  q_{\rm naive} = \frac{\sum_{\mu}\sum_{x}b(x\, |\, \hat{\mu})}{\sum_{\mu}\sum_{x}c(x\, |\, \hat{\mu})}\, .
\end{equation}
The associated physical quantity is the rate per unit length, $\lambda_{\rm
  naive}=q_{\rm naive}/a$. The calculated rate parameters from both methods are
shown in Table~\ref{tab:rate}. We observe that with both measures the physical
branching rate increases as the physical pion mass is approached.  We emphasise,
only $q$ contains the detailed information on the path geometry.

\begin{table*}[t]
  \caption{The naive and fitted branching rates, $q_{\rm naive}$ and
    $q$, and their physical counterparts $\lambda_{\rm naive}$ and $\lambda$
    obtained through the methods described in the text. The fit parameter
    $\beta$ is also presented. Only $q$ and $\lambda$ are associated with a
    constant branching probability. }
  \label{tab:rate}
  \begin{ruledtabular}
    \begin{tabular}{ldddddddd}
      \input{./include/rate_table.tex}
    \end{tabular}
  \end{ruledtabular}
\end{table*}


The difference between the fitted and naive rates is an interesting finding.
The naive rate will include the short-range non-exponential behaviour,
inconsistent with a constant branching rate.
At larger separations, vortex branching follows a constant rate.
However, there are clearly short-range effects that result in clustering of
branching points, which in turn necessitates the more sophisticated approach
detailed above for $q$.
These clustering effects appear to be amplified upon introduction of dynamical
fermions. Whether this clustering radius is a physical effect or the result of
finite lattice-spacing effects is an interesting avenue for future study.

It should be noted that whilst the distributions shown in
Fig.~\ref{fig:DistHistAll} take into account all primary and secondary clusters,
the results are minimally affected if the secondary clusters are removed due to
the vast majority of branching points belonging to the primary cluster.

An interesting correlation we observe is that the ratio between the pure gauge
and dynamical branching rates is similar to the corresponding ratio of the
vortex-only string tensions calculated in Ref.~\cite{Biddle:2022zgw}. The vortex
density is naturally correlated with the branching rate. In SU(2) at least, it
has been shown through simple combinatoric arguments that the Wilson loop area
law and hence the string tension can be related to the density of percolating
random vortices~\cite{Engelhardt:1998wu}. It seems reasonable to infer then that
the correlation we observe between the branching rate and string tension ratios
is not simply a coincidence but a reflection of the differing structure of the
vortex fields in the pure gauge and dynamical sectors.

\section{Conclusion}\label{sec:Conclusion}

In this work we have explored of the impact of dynamical fermions on the
centre-vortex structure of the vacuum ground-state fields.

Examining the bulk properties of the original gauge fields, we find that
dynamical fermions lead to greater off-diagonal strength in the lattice gauge
links. The presence of dynamical fermions gives rise to an increased abundance
of centre vortices and branching points, as reflected by the increasing vortex
and branching point densities as the physical pion mass is approached.

We construct cluster identification algorithms to identify independent vortex
clusters and use this identification to construct visualisations of the vortex
vacuum. These reveal that the vacuum is dominated by a single percolating
cluster. Our results show that dynamical fermions lead to an abundance of
smaller clusters as compared to their pure-gauge counterparts.

We employ a novel method of reducing vortex clusters to directed graphs, with
vertices defined by branching points and edges connecting them weighted by the
number of vortex links. Using this construction, we render the graphs to
illustrate the radical change in the number of vortices and branching points
after the introduction of dynamical fermions. We define
a measure of branching point separation, and observe that the distribution of
separations follows an approximate geometric distribution. We estimate the
rate of this distribution and find that there is a tendency for branching points
to cluster at small separations.

Understanding the role of dynamical quarks in the QCD vacuum continues to be an
interesting area of study. The effect of matter fields on the vacuum phase
structure has been explored elsewhere within the gauge-Higgs
model~\cite{Greensite:2017ajx,Greensite:2018mhh,Greensite:2020nhg,Greensite:2021fyi}.
The extension of these ideas to QCD may shed further light on the nature of
confinement. In particular, investigations that further our understanding of
string breaking in terms of QCD vacuum structure is desirable.

The findings of this paper illustrate the substantial impact dynamical fermions
have on the geometry and structure of the centre vortex vacuum. These results
add to the growing body of evidence~\cite{Biddle:2022acd,Biddle:2022zgw}
for the effect of dynamical fermions on centre vortices as compared to the
well-established pure gauge sector. The relationship between the
vortex geometry analysed here and the shift in observable behaviour is still a
subject of great interest. Future work is also intended to explore how this
geometry changes in the finite temperature regime.

\section*{Acknowledgements}
We thank the PACS-CS Collaboration for making their 2 +1 flavour configurations
available via the International Lattice Data Grid (ILDG). This research was
undertaken with the assistance of resources from the National Computational
Infrastructure (NCI), provided through the National Computational Merit
Allocation Scheme and supported by the Australian Government through Grant No.
LE190100021 via the University of Adelaide Partner Share. This research is
supported by Australian Research Council through Grants No. DP190102215 and
DP210103706. WK is supported by the Pawsey Supercomputing Centre through the
Pawsey Centre for Extreme Scale Readiness (PaCER) program.

\bibliography{vortex_structure}

%
\newpage
\widetext

\renewcommand{\thefigure}{S-\arabic{figure}}

\centerline{\Large\bf Centre vortex structure in the presence of dynamical fermions}
\vspace{3pt}
\centerline{\bf Supplemental Material}
\vspace{12pt}
\centerline{James C. Biddle, Waseem Kamleh, and Derek B. Leinweber}
\vspace{12pt}
\centerline{\it Centre for the Subatomic Structure of Matter, Department of Physics, The University of Adelaide, SA}
\centerline{\it 5005, Australia}
\vspace{12pt}

This supplementary document contains interactive 3D models embedded in the text, complementary to
the static images presented in the main text. To interact with these models, it is necessary to
open this document in Adobe Reader or Adobe Acrobat (requires version 9 or newer). Linux users may
install Adobe Acroread version 9.4.1, or use a Windows emulator such as
\href{https://www.playonlinux.com/en/}{PlayOnLinux}. 3D content must be enabled for the interactive
content to be available, and for proper rendering it is necessary to enable double-sided rendering
in the preferences menu. 

To activate the models, simply click on the image. To rotate the model, click and hold the left
mouse button and move the mouse. Use the scroll wheel or shift-click to zoom. Some pre-set views of
the model are also provided to highlight areas of interest. These can be accessed by right clicking
and using the ``Views'' menu. To reset the model back to its original orientation and zoom, press
the ‘home’ icon in the toolbar or change the view to ‘Default view’.

\clearpage

%
\begin{figure}[p]
\centering
\null\hspace{-0.3cm}
\includemedia[
        noplaybutton,
	3Dtoolbar,
	3Dmenu,
	3Dviews=U3D/Primary-Secondary-PG-24.vws,
	3Dcoo  = 16 16 32, 
        3Dc2c=0.245114266872406 0.8673180341720581 0.43321868777275085,
	3Droo  = 110.0,    
	3Droll =-98.1,     
	3Dlights=Default,  
	width=\textwidth,  
]{\includegraphics{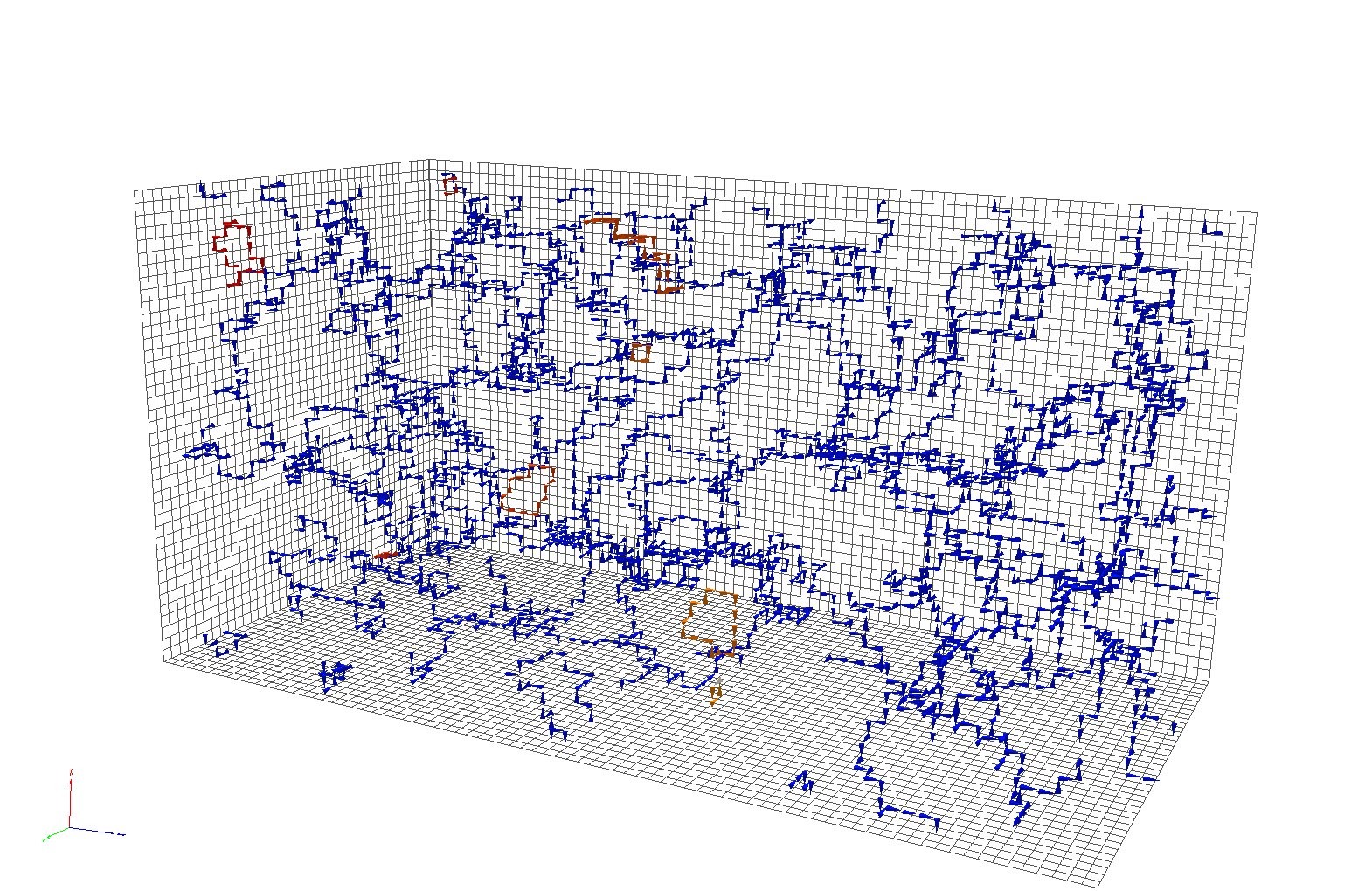}}{U3D/Primary-Secondary-PG-24.u3d}
\caption{ The centre vortex structure of a ground-state vacuum field configuration in pure SU(3)
  gauge theory. The flow of $+1$ centre charge through the gauge field is illustrated by the jets
  (see main text for a description of the plotting conventions). Blue jets are used to illustrate
  the primary percolating vortex cluster, while other colours illustrate the secondary
  clusters. ({\sl Click to activate.})
\label{sup-fig:Primary-Secondary-PG-24.u3d} }

  \centering
  \includemedia[
        noplaybutton,
	3Dtoolbar,
	3Dmenu,
	3Dviews=U3D/Primary-Secondary-DF-05.vws,
	3Dcoo  = 16 16 32, 
        3Dc2c=0.245114266872406 0.8673180341720581 0.43321868777275085,
	3Droo  = 110.0,    
	3Droll =-98.1,     
	3Dlights=Default,  
	width=\textwidth,  
]{\includegraphics{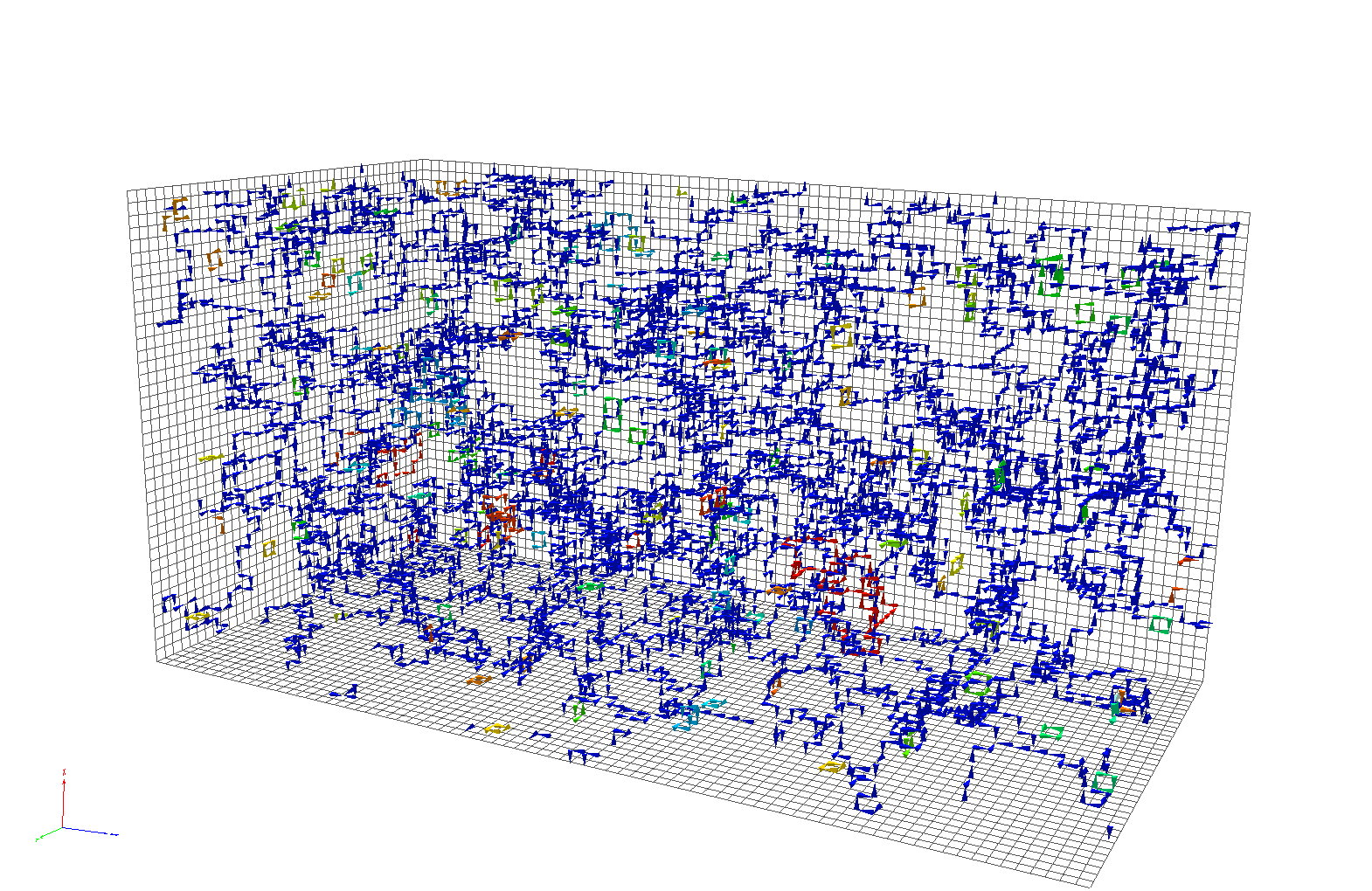}}{U3D/Primary-Secondary-DF-05.u3d}
\caption{ The centre-vortex structure of a ground-state vacuum field configuration in dynamical 2+1
  flavour QCD with $m_{\pi} = 156 ~ \si{MeV}$.  Symbols are as described in
  Fig.~\ref{sup-fig:Primary-Secondary-PG-24.u3d}. ({\sl Click to activate.})
\label{sup-fig:Primary-Secondary-DF-05.u3d}}
\end{figure}

\begin{figure}[p]
\centering
\includemedia[
        noplaybutton,
	3Dtoolbar,
	3Dmenu,
	3Dviews=U3D/Secondary-PG-24.vws,
	3Dcoo  = 16 16 32, 
        3Dc2c=0.245114266872406 0.8673180341720581 0.43321868777275085,
	3Droo  = 110.0,    
	3Droll =-98.1,     
	3Dlights=Default,  
	width=\textwidth,  
]{\includegraphics{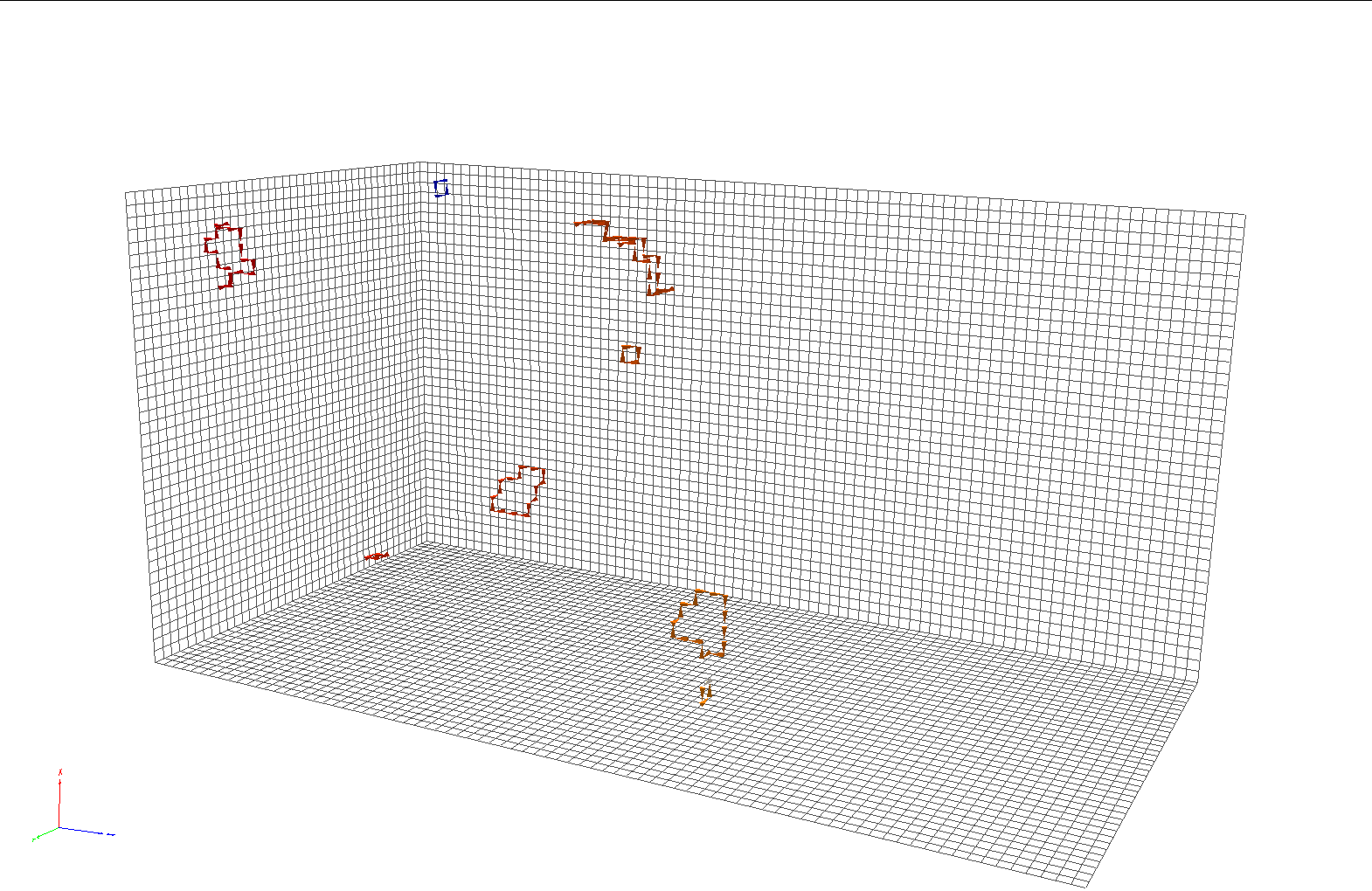}}{U3D/Secondary-PG-24.u3d}
\caption{ The centre-vortex structure of the secondary loops identified from the pure-gauge
  configuration shown in Fig.~\ref{sup-fig:Primary-Secondary-PG-24.u3d}. ({\sl Click to activate.})
\label{sup-fig:Secondary-PG-24.u3d} }

  \centering
\includemedia[
        noplaybutton,
	3Dtoolbar,
	3Dmenu,
	3Dviews=U3D/Secondary-DF-05.vws,
	3Dcoo  = 16 16 32, 
        3Dc2c=0.245114266872406 0.8673180341720581 0.43321868777275085,
	3Droo  = 110.0,    
	3Droll =-98.1,     
	3Dlights=Default,  
	width=\textwidth,  
]{\includegraphics{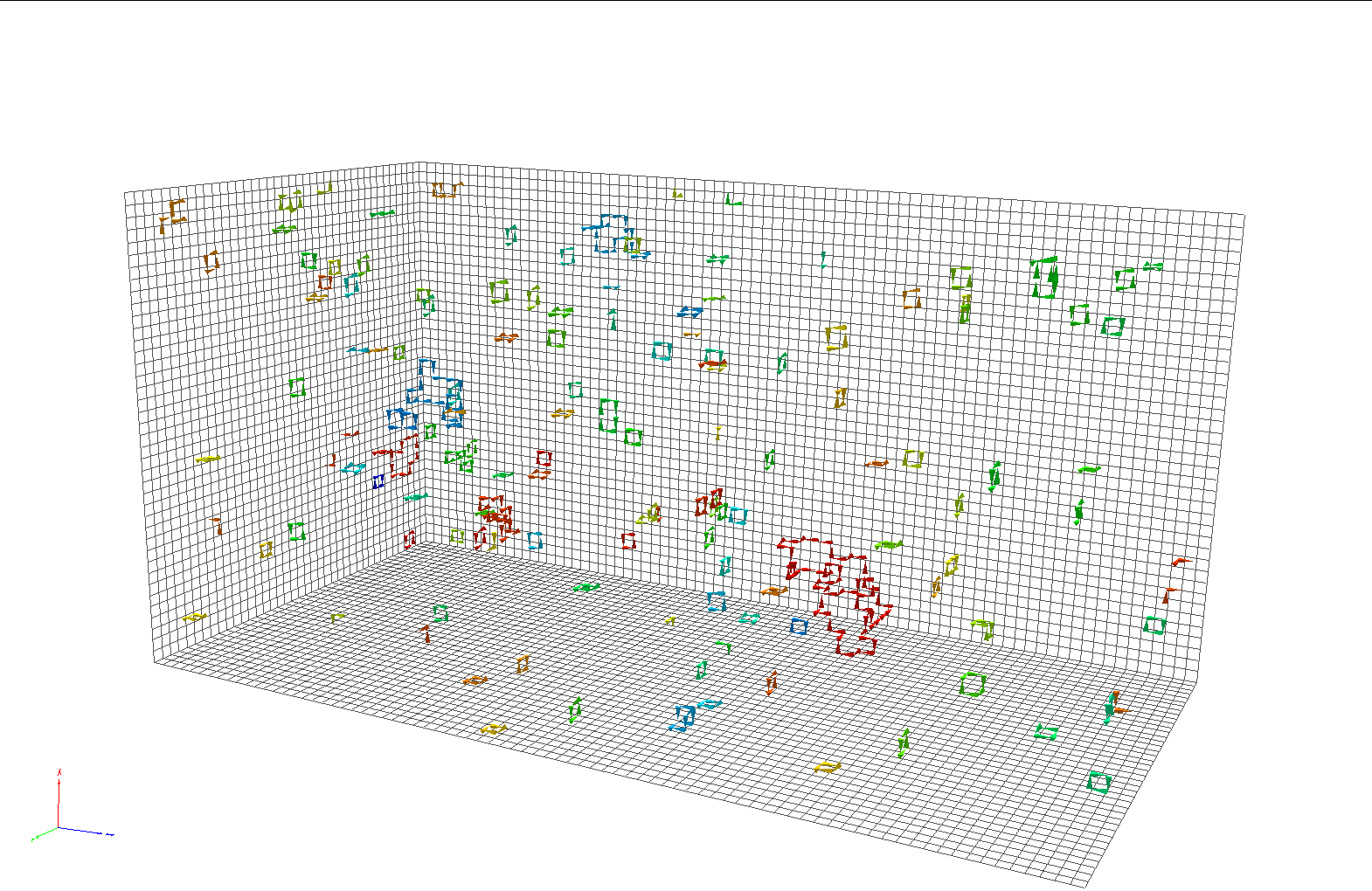}}{U3D/Secondary-DF-05.u3d}
\caption{ The centre-vortex structure of the secondary loops identified from the dynamical-fermion
  configuration shown in Fig.~\ref{sup-fig:Primary-Secondary-DF-05.u3d}. ({\sl Click to activate.})
\label{sup-fig:Secondary-DF-05.u3d}}
\end{figure}

\end{document}

%% file: include/loop_table.tex
 & $\hat{t}$ & $\hat{x},\hat{y},\hat{z}$ \\
\midrule
\multicolumn{1}{l}{Pure gauge}\\
\midrule
$N_{\rm slice}$ & 1673(3) & 3347(6) \\
$N_{\rm primary}$ & 1638(3) & 3277(6) \\
$N_{\rm secondary}$ & 7.32(5) & 7.40(3) \\
\midrule
\multicolumn{1}{l}{$701~\si{MeV}$}\\
\midrule
$N_{\rm slice}$ & 3651(4) & 7302(8) \\
$N_{\rm primary}$ & 3366(4) & 6731(8) \\
$N_{\rm secondary}$ & 5.047(5) & 5.057(3) \\
\midrule
\multicolumn{1}{l}{$156~\si{MeV}$}\\
\midrule
$N_{\rm slice}$ & 3227(4) & 6452(8) \\
$N_{\rm primary}$ & 2964(4) & 5926(9) \\
$N_{\rm secondary}$ & 5.011(5) & 5.018(3) 

%% file: include/vortex_density_table.tex
         &                  & $\rho_{\rm vortex}$ &                  & $\rho_{\rm branch}$ \\
Ensemble & $P_{\rm vortex}$ & $(\si{fm^{-2}})$    & $P_{\rm branch}$ & $(\si{fm^{-3}})$    \\
\midrule
Pure gauge     & 0.01702(3) & 1.702(3) & 0.00249(1)  & 2.49(1) \\
$701~\si{MeV}$ & 0.03714(4) & 3.556(4) & 0.00897(1)  & 8.41(1) \\
$156~\si{MeV}$ & 0.03282(4) & 3.770(5) & 0.00753(1)  & 9.27(2) 

%% file: include/graph_stats_table.tex
Ensemble & \mc{\quad $d$} & \mc{\quad\qquad $\Delta~(\si{fm})$}
&$n_{\rm edges}$ & \mc{\qquad $\rho_{\rm edges}~(\si{fm^{-3}})$} & \mc{\qquad\qquad$n_{\rm edges}/n_{\rm nodes}$} \\
\midrule
Pure gauge          & 13.55(2) & 1.355(2)  & 238(1) & 4.14(1)  & 1.53849(8) \\
$701~\si{MeV}$      & 7.691(4) & 0.7860(4) & 970(1) & 15.84(2) & 1.58667(6) \\
$156~\si{MeV}$      & 8.082(5) & 0.7541(5) & 807(1) & 17.32(3) & 1.58332(7) 

%% file: include/rate_table.tex
Ensemble & \mc{\quad\qquad$q_{\rm naive}$} & \mc{\quad\qquad$\lambda_{\rm naive}~(\si{fm^{-1}})$} & \mc{\qquad$q$} & \mc{\quad\qquad$\lambda~(\si{fm^{-1}})$} & \mc{\qquad$\beta$} \\
\midrule
Pure gauge     & 0.05010(6) & 0.5010(6) & 0.0690(2) & 0.690(2) & 0.0715(2) \\
$701~\si{MeV}$ & 0.08526(5) & 0.8342(5) & 0.1005(3) & 0.984(3) & 0.1059(3) \\
$156~\si{MeV}$ & 0.08062(6) & 0.8641(7) & 0.0952(2) & 1.020(3) & 0.1000(3)